\documentclass[useAMS,usenatbib]{mn2e}

\usepackage{graphicx, amsmath}

\newcommand{\RV}{V_\mathrm{los}}

\newcommand{\kms}{\ \mathrm{kms}^{-1}}
\newcommand{\teff}{T_\mathrm{eff}}

\newcommand{\VPHI}{V_\phi}
\newcommand{\vlos}{V_\mathrm{los}}
\newcommand{\VR}{V_R}
\newcommand{\VZ}{V_Z}
\newcommand{\pc}{\textrm{pc}}
\newcommand{\mua}{\mu_\alpha}
\newcommand{\mud}{\mu_\delta}
\newcommand{\kpc}{\mathrm{kpc}}
\newcommand{\masyr}{\mathrm{masyr}^{-1}}
\newcommand{\Gyr}{\mathrm{Gyr}}

\begin{document}
\title[The wobbly Galaxy]{The wobbly Galaxy: kinematics north and south with RAVE red clump giants}

\author[M. E. K Williams et al.]{M. E. K. Williams$^{1}$\thanks{E-mail: mary@aip.de}, M. Steinmetz$^{1}$, J. Binney$^{2}$, A. Siebert$^{3}$, H. Enke$^{1}$, B. Famaey$^3$,
\newauthor I. Minchev$^{1}$, R. S. de Jong$^{1}$, C. Boeche$^{4}$, K. C. Freeman$^{5}$, O. Bienaym{\'e}$^{3}$,
\newauthor J. Bland-Hawthorn$^{6}$, B. K. Gibson$^{7}$, G. F. Gilmore$^{8}$, E. K. Grebel$^{4}$, A. Helmi$^{9}$,
\newauthor G. Kordopatis$^{8}$, U. Munari$^{10}$, J. F. Navarro$^{11}$, Q. A. Parker$^{12}$, W. Reid$^{12}$, 
\newauthor G. M. Seabroke$^{13}$, S. Sharma$^{6}$, A. Siviero$^{14, 1}$, F. G. Watson$^{15}$, R. F. G. Wyse$^{16}$, \newauthor T. Zwitter$^{17}$\\
$^{1}${Leibniz Institut f\" ur Astrophysik Potsdam (AIP), An der Sterwarte 16, D-14482 Potsdam, Germany}\\
$^{2}${Rudolf Pierls Center for Theoretical Physics, University of Oxford, 1 Keble Road, Oxford OX1 3NP, UK}\\
$^{3}${Observatoire astronomique de Strasbourg, Universit\'e de Strasbourg, CNRS, UMR 7550, Strasbourg, France}\\
$^{4}${Astronomisches Rechen-Institut, Zentrum f\"ur Astronomie der Universit\"at Heidelberg, D-69120 Heidelberg, Germany}\\
$^{5}${RSAA Australian National University, Mount Stromlo Observatory, Cotter Road, Weston Creek, Canberra, ACT
72611, Australia}\\
$^{6}${Sydney Institute for Astronomy, School of Physics, University of Sydney, NSW 2006, Australia}\\
$^{7}${Jeremiah Horrocks Institute for Astrophysics \& Super-computing, University of Central Lancashire, Preston, UK}\\
$^{8}${Institute of Astronomy, University of Cambridge, Madingley Road, Cambridge CB3 0HA, UK}\\
$^{9}${Kapteyn Astronomical Institute, University of Groningen, Postbus 800, 9700 AV Groningen, Netherlands}\\
$^{10}${INAF - Astronomical Observatory of Padova, 36012 Asiago (VI), Italy}\\
$^{11}${Senior CIfAR Fellow, University of Victoria, P.O. Box 3055, Station CSC, Victoria, BC V8W 3P6, Canada}\\
$^{12}${Macquarie University, Sydney, NSW 2109, Australia}\\
$^{13}${Mullard Space Science Laboratory, University College London, Holmbury St Mary, Dorking, RH5 6NT, UK}\\
$^{14}${Department of Physics and Astronomy ``Galileo Galilei", Padova University, Vicolo dellÕOsservatorio 2, I-35122 Padova, Italy}\\
$^{15}${Anglo-Australian Observatory, P.O. Box 296, Epping, NSW 1710, Australia}\\
$^{16}${Johns Hopkins University, 3400 N Charles Street, Baltimore, MD 21218, USA}\\
$^{17}${Faculty of Mathematics and Physics, University of Ljubljana, Jadranska 19, Ljubljana, Slovenia}}

\date{Accepted August 12, 2013 }

\pagerange{1--22} \pubyear{2013}

\maketitle

\begin{abstract}
The RAVE survey, combined with proper motions and distance estimates, can be used to study in detail stellar kinematics in the extended solar neighbourhood (solar suburb). Using $72,365$ red clump stars, we examine the mean velocity components in 3D between $6<R<10\,\kpc$ and $-2<Z<2\,\kpc$, concentrating on North-South differences. Simple parametric fits to the $(R,\:Z)$ trends for $\VPHI$ and the velocity dispersions are presented. We confirm the recently discovered gradient in mean Galactocentric radial velocity, $\VR$, finding that the gradient is marked below the plane ($\delta \langle \VR \rangle/ \delta R=-8\,\kms/\kpc$ for $Z<0$, vanishing to zero above the plane), with a $Z$ gradient thus also present. The vertical velocity, $\VZ$, also shows clear, large-amplitude ($|\VZ|=17\,\kms$) structure, with indications of a rarefaction-compression pattern, suggestive of wave-like behaviour. We perform a rigorous error analysis, tracing sources of both systematic and random errors. We confirm the North-South differences in $\VR$ and $\VZ$ along the line-of-sight, with the $\VR$ estimated independent of the proper motions. The complex three-dimensional structure of velocity space presents challenges for future modelling of the Galactic disk, with the Galactic bar, spiral arms and excitation of wave-like structures all probably playing a role. 
\end{abstract}


\section{Introduction}
\label{sec1}

The more we learn about our Galaxy, the Milky Way, the more we find evidence
that it is not in a quiescent, stable state. Rather, the emerging picture is of a Galaxy in flux, evolving under the forces of internal and external
interactions. Within the halo, there is debris left by accretion
events, the most significant being the Sagittarius stream \citep{Majewski2003}
at a mean distance of $d\sim20-40\,\kpc$  associated with the Sagittarius dwarf galaxy \citep{Ibata1994}. In the intersection of halo and
disk there are also indications of accretion debris, such as the Aquarius Stream
($0.5<d<10\,\kpc$, \citealt{Williams2011}) as well as local halo streams such as the Helmi stream
($d<2.5\,\kpc$, \citealt{Helmi1999}). While within the disk itself
there is evidence of large-scale stellar over-densities: outward from the Sun at a Galactocentric distance of $R=18-20\,\kpc$ there is the Monoceros Stream
(\citealt{Newberg2002}, \citealt{Yanny2003}), inward from the Sun ($d\sim2\,\kpc$) there is the Hercules thick disk cloud
(\citealt{Larsen1996}, \citealt{Parker2003}, \citealt{Parker2004},
\citealt{Larsen2008}).  
The origin of all such structures is not yet fully understood: possible
agents for the Monoceros ring are tidal debris (see e.g.,
\citealt{Penarrubia2005}), the excitation of the disk caused by accretion events
(e.g., \citealt{Kazantzidis2008}) or the Galactic warp
(\citealt{Momany2006}), while \citealt{Humphreys2011} conclude that the Hercules
cloud is most likely due to a dynamical interaction of the thick disk with the Galaxy's bar. 

Velocity space also exhibits structural complexity. In the solar neighbourhood
various structures are observed in the distribution of stars in the $UV$
plane,  which differs significantly from a
smooth Schwarzschild distribution \citep{Dehnen1998}. These features are likely created by the combined effects of the Galactic bar (\citealt{Raboud1998}, \citealt{Dehnen1999}, \citealt{Minchev2010},
\citealt{Famaey2007}) and spiral arms (\citealt {deSimone2004, Quillen2005}, \citealt{Antoja2009}) or both (\citealt{Chakrabarty2007}, \citealt{Quillen2011}). Dissolving open clusters (\citealt{Skuljan1997}, \citealt{deSilva2007}) can also contribute to structure in the $UV$ plane. Finally, velocity streams in the $UV$ plane can be explained by the
perturbative effect on the disk by recent merger events (\citealt{Minchev2009}, \citealt{Gomez2012}).  

Venturing beyond the solar neighbourhood into the solar suburb, \citealt{Antoja2012} used data from the RAdial Velocity
Experiment (RAVE) \citep{Steinmetz2006} to show how these resonant features could be traced far from the Sun, up to $1\,\kpc$ from the Sun in the direction of anti-rotation and $0.7\,\kpc$ below the Galactic plane. Additionally, \citealt{Siebert2011} (hereafter S11), showed based on RAVE data that the components of
stellar velocities in the direction of the Galactic centre, $\VR$, is
systematically non-zero and has a non-zero gradient $\delta \langle \VR
\rangle/ \delta R<-3\,\kms/\kpc$. A similar effect was seen in the analysis of 4400 RAVE red clump giants by \citealt{Casetti2011}. The cause of
this streaming motion has been variously ascribed to the bar, spiral arms,
and the warp in conjunction with a triaxial dark-matter halo, or a
combination of all three. Recently, \citealt{Siebert2012} have used
density-wave models to explore the possibility that spiral arms cause the
radial streaming, and were able to reproduce the gradient with a
two-dimensional model in which the Sun lies close to the 4:1 ultra-harmonic
resonance of two-arm spiral, in agreement with \citealt{Quillen2005} and \citealt{Pompeia2011}. While structure in the $UV$ plane smears out with the increase of
sample depth (at $d>250\, \mathrm{pc}$), Gomez et al. (2012a,b) showed that
energy-angular momentum space preserves structure associated with 
 ``ringing" of the disk in response to a minor merger event for distances around the Sun as large as $3\,\kpc$, consistent with the SEGUE and RAVE coverage.

In this paper we examine the kinematics of the red clump giants in the
three-dimensional volume surveyed by RAVE, focusing particularly on
differences between the northern and southern sides of the plane. The full space velocities are calculated from RAVE line-of-sight velocities ($\RV$), literature proper motions and photometric distances from the red clump. We have sufficient stars from the RAVE internal data
set for the stellar velocity field to be
explored within the region $6<R<10\,\kpc$ and
$-2<Z<2\,\kpc$. While studies such as \citealt{Pasetto2012a, Pasetto2012b}, distinguish between thick- and thin-disk stars, we do not. Distinguishing between the thin, thick disk and halo require either kinematic or chemical criteria, each of which have their individual issues and challenges, which we wish to avoid. Rather, our aim here is to describe the overall velocity structure of the solar neighbourhood in a pure observational/phenomenological sense. The interpretation of this based on thin or thick-disk behaviour is then secondary.

In S11, $\VR$ as a function of $(X,\ Y)$ was examined. Here we extend this analysis to investigate particularly the $Z$ dependence of $\VR$, as well as the other two velocity components, $\VPHI$ and $\VZ$, with a simple parametric fit for $\VPHI$ presented. We find that the average values of $\VR$ and $\VZ$ show a strong dependence on three-dimensional position. We compare these results to predictions of a steady-state, axisymmetric Galaxy from the \textsc{Galaxia} \citep{Sharma2011} code. To visualise the 3D behaviour better, various projections are used. Further, the detection by S11 of the gradient in $\VR$ using only line-of-sight velocities is re-examined in light of the results seen in 3D, extending the analysis to include $\VZ$. As an aside, parametric fits to the velocity dispersions are presented.

The paper also includes a thorough investigation into the effects of systematic and measurement errors,
which almost everywhere dominate Poisson noise. The assumptions used to calculate photometric distances from the red clump are examined in some detail, where we model the clump using \textsc{Galaxia}. The effects of
using alternative sources of proper motion are shown in comparison with the main results, as we found this to be one of the largest systematic error sources.

The paper is organised as follows. In Section \ref{sec2} we discuss our overall approach for the kinematic mapping, the co-ordinate systems used and initial selection from RAVE. In Section \ref{sec3} we present a detailed investigation into the use of the red clump stars as a distance indicator, where we introduce a \textsc{Galaxia} model of the red clump to model the distance systematics. In Section \ref{sec4} we present our error analysis, investigating systematic error sources first before discussing measurement errors, Poisson noise and the final cuts to the data using the error values. In Section \ref{sec5} we examine the 3D spatial distribution of our data before going on in Section \ref{sec6} to investigate the variation for $\VR,\,\VPHI$ and $\VZ$ over this space. Also presented are the variations in velocity dispersions with $(R,\,Z)$ and a simple functional fit to these trends. In Section \ref{sec7} we use the line-of-sight method to investigate the $\VR$ and $\VZ$ trends towards and away from the Galactic centre, while finally, Section \ref{sec8} contains a summary of our results.

\section{Red Clump kinematics with RAVE data}
\label{sec2}

\subsection{Co-ordinate systems and Galactic parameters}
\label{sec2.2}

As we are working at large distances from the solar position, we use cylindrical co-ordinates for the most part, with $\VR,\,\VPHI$ and $\VZ$ defined as positive with increasing $R$, $\phi$ and $Z$, with the latter towards the North Galactic Pole (NGP). We also use a right-handed Cartesian co-ordinate system with $X$ increasing towards the Galactic centre, $Y$ in the direction of rotation and $Z$ again positive towards the NGP. The Galactic centre is located at $(X,Y,Z)=(R_\odot,\,0,\,0)$. Space velocities, $UVW$, are defined in this system, with $U$ positive towards the Galactic centre. 

To aid comparison with the predictions of \textsc{Galaxia} (see Section \ref{sec2.4}), we use mostly the
parameter values that were used to make the model for which \textsc{Galaxia} makes predictions. Thus, the motion of the Sun with respect to the LSR is
taken from \citealt{Schoenrich2010}, namely $U_\odot=11.1,\
V_\odot=12.24,W_\odot=7.25\,\kms$. The LSR is assumed to be on a circular
orbit with circular velocity $V_\mathrm{circ}=226.84\,\kms$. Finally, we take $R_\odot=8\,\kpc$. The only deviation from
\textsc{Galaxia}'s values is that they assumed that the Sun is located at $Z=+15\,\pc$, while we assume $Z=0\,\pc$.

S11 explored the variation of the observed $\VR$ gradient on the values of $V_\mathrm{circ}$ and $R_\odot$, finding that changing these parameters between variously accepted values could reduce but not eliminate the observed gradient in $\VR$. We do not explore this in detail in this paper but note that the trends that we observe are similarly affected by the Galactic parameters; amplitudes are changed but the qualitative trends remain the same. 

\subsection{RAVE data}
\label{sec2.3}

The wide-field RAVE (RAdial Velocity Experiment) survey measures primarily line-of-sight velocities and additionally stellar parameters, metallicities and abundance ratios of stars in the solar neighbourhood \citep{Steinmetz2006, Zwitter2008, Siebert2011, Boeche2011}.  RAVE's input catalog is magnitude limited ($8<I<13$) and thus creates a sample with no kinematic biases. To the end of 2012 RAVE had collected more than 550,000 spectra with a median error of $1.2\,\kms$ \citep{Siebert2011b}.

\begin{table*}
\begin{footnotesize}
\begin{tabular}{lllll}
\hline
Version&$M_{K_s}$&Source\\
 \hline
A&$-1.65$&Observation: \citealt{Alves2000}, \citealt{Grocholski2002}\\
B&$-1.54$&Observation: \citealt{Groenewegen2008}\\
C&$-1.64+0.0625 | Z(\kpc) | $&Theory: \citealt{Salaris2002} compared to the RAVE population\\
\end{tabular}
\caption{The normalisations used for the red clump $M_{K_s}$ magnitudes}
\label{tab1}
\end{footnotesize}
\end{table*}

We use the internal release of RAVE from October 2011 which contains 434,807 RVs and utilizes the revised stellar parameter determination (see the DR3 paper, \citealt{Siebert2011b} for details). We applied a series of cuts to the data. Firstly, those stars flagged by the \citealt{Matijevic2012} automated spectra classification code as having peculiar spectra were excluded. This removes most spectroscopic binaries, chromospherically active and carbon stars, spectra with continuum abnormalities and other unusual spectra. Further cuts restricted our sample to stars with signal-to-noise ratio $STN>20$ (STN is calculated from the observed spectrum alone with residuals from smoothing, see Section 2.2 of the DR3), \citealt{Tonry1979} cross-correlation coefficient $R>5$, $|\mu_\alpha,\ \mu_\delta|<400\,\masyr$, $e\mu_\alpha,\ e\mu_\delta < 20\,\masyr$, $\mathrm{RV}<600\,\kms$ (see Section \ref{sec4.4}) and stars whose SpectraQualityFLAG is null. Where there are repeat observations, we randomly select one observation for each star. With this cleaning, the data set has 293,273 unique stars with stellar parameters from which to select the red clump. 

In some fields at $|b| < 25^\circ $ the RAVE selection function includes a
colour cut $J-K > 0.5$ with the object of favouring giants. We imposed this
colour cut throughout this region to facilitate the comparison to predictions by \textsc{Galaxia}. The selection of red clump stars is unaffected by this cut.
Additional data cuts and selections were performed later in the analysis and
are broadly a) the selection of red-clump giants (Section \ref{sec3.3}), b)
removal of stars with large extinction/reddening (Section \ref{sec4.1.4}),
and removal of data bins with large errors in measurements of mean velocity
(Section \ref{sec4.4}). 

As a sample with alternative distance determinations, we also used an internal release with stellar parameters produced by the pipeline that was used for the third Data Release (VDR3) and the method of \citealt{Zwitter2008} from September 2011 with 334,409 objects. We cleaned the sample as above, leaving us with 301,298 stars. 

A newer analysis of RAVE data is presented in the 4th RAVE data release, Kordopatis et al. (in preparation). This applies an updated version of the \citealt{Kordopatis2011} pipeline to RAVE spectra. Selecting red clump stars using the stellar parameters from this pipeline does not produce significantly different results to those selected using the VDR3 pipeline above; the conclusions of this paper are unaffected by the pipeline used for the red-clump selection. 

\subsection{The \textsc{Galaxia} model}
\label{sec2.4}

In \citealt{Williams2011} we introduced the use of the Galaxy modelling code \textsc{Galaxia} \citep{Sharma2011} to investigate the statistical significance of the new Aquarius stream found with RAVE. In this study we use \textsc{Galaxia} both to provide predictions with which to compare our results, and to investigate the effects of contamination of the red-clump sample by stars that are making their first ascent of the giant branch (see Section \ref{sec4.1.1}). \textsc{Galaxia} enables us to disentangle real effects from artifacts of our methodology, and further to understand the population we are examining.

Based on the Besan\c con Galaxy model, the \textsc{Galaxia} code creates a synthetic catalog of stars for a given model of the Milky Way. 
It offers several improvements over the Besan\c con model which increase its utility in modelling a large-scale survey like RAVE, the most significant of which is the ability to create a continuous distribution across the sky instead of discrete sample points. The elements of the Galactic model are a star formation rate, age-velocity relation, initial mass function and density profiles of the Galactic components (thin and thick disk, smooth spheroid, bulge and dark halo). The parameters and functional forms of these components are summarized in Table 1 of \citealt{Sharma2011}.

We follow a similar methodology to that described in \citealt{Williams2011} to generate a \textsc{Galaxia} model of the RAVE sample, from which we select a red clump sample following the selection criteria applied to the real data. A full catalog was generated over the area specified by $0<l<360$, $\delta<2^\circ$ and $9<I<13$, with no under-sampling. As described in Section \ref{sec3.2}, the RAVE data were then divided into three signal-to-noise regimes and a sample drawn from the \textsc{Galaxia} model so that, for each regime, the $I$-band distribution was matched to that of the RAVE sample in $5^\circ \times
5^\circ$ squares. The \textsc{Galaxia} $I$-band was generated after correcting for extinction.

\section{The Red Clump}
\label{sec3}

The helium-burning intermediate-age red clump has long been seen as a promising standard candle \citep{Cannon1970}, with its ease of identification on the HR diagram. In recent years there has been a renewed interest in the red clump for distance determination, e.g., \citealt{Pietrzynski2003} studied the clump in the Local Group, down to the metal-poor Fornax dwarf galaxy. Here we investigate the use, selection and modelling of this population in solar-suburb RAVE data. 

\subsection{The Red Clump $K$-band magnitude}
\label{sec3.1}

The $K$-band magnitude of the red clump, while being relatively unaffected by
extinction, has also been shown observationally to be only weakly dependent
on metallicity and age \citep{Pietrzynski2003, Alves2000}, so a single
magnitude is usually assigned for all stars in the red clump.  Studies
such as \citealt{Grocholski2002} and \citealt{vanHelshoecht2007} have shown that
there is some dependence on metallicity and age which can be accounted for by
the theoretical model of \citealt{Salaris2002}. 

If models correctly predict the systematic dependence of $M_K$ of the red clump on metallicity
and age, this would have implications for our study of kinematics in the
solar suburb, for with increasing distance above the plane the metallicity
will decrease and the population become older.  From \citealt{Burnett2011} we estimate
that the population means change from $\mathrm{[M/H]}\sim0$,
$\mathrm{Age}=4\,\mathrm{Gyr}$ at $Z=0\,\kpc$ to $\mathrm{[M/H]}\sim-0.6$,
$\mathrm{Age}=10\,\mathrm{Gyr}$ at $Z=|4|\,\kpc$.  According to
\citealt{Salaris2002}, these age and metallicity changes will change the
$K$-band absolute magnitude from $M_{K_s}\textrm{(RC)}=-1.64$\footnote{Here $K$ is
used to denote $K$-band magnitudes in the \citealt{Bessell1989} system, while
$K_s$ is used to denote 2MASS values. The relations from
\citealt{Carpenter2003} were used to convert between the two photometric
systems. Note though that $J-K_S$ is shortened to $J-K$ in Section \ref{sec3.2} and beyond.} at $Z=0\,\kpc$ to $M_{K_s}\textrm{(RC)}=-1.39$ at $|Z|=4\,\kpc$. Hence,
the distances to stars at higher $Z$ will be systematically underestimated by
$\sim10\:\mathrm{per\:cent}$. 

Furthermore, there is uncertainty regarding the average value of $M_K$.
\citealt{Alves2000} gives $M_K\textrm{(RC)}=-1.61\pm0.03$ for local red clump
giants with Hipparcos distances and metallicities between $-0.5\le
\mathrm{[Fe/H]}\le0.0$. In the 2MASS system the corresponding absolute
magnitude is $M_{K_s}\textrm{(RC)}=-1.65\pm0.03$.  \citealt{Grocholski2002}
derives a similar value of $M_K\textrm{(RC)}=-1.61\pm0.04$ for $-0.5\le
\mathrm{[Fe/H]}\le0.0$, $1.6\,\Gyr \le \mathrm{Age}\le8\,\Gyr$ from 2MASS
data of open clusters.  Extending the sample of clusters,
\citealt{vanHelshoecht2007} derived $M_K\textrm{(RC)}=-1.57\pm0.05$ for
$-0.5\le \mathrm{[Fe/H]}\le0.4$, $0.3\,\Gyr \le \mathrm{Age}\le8\,\Gyr$. Then
using the new \citealt{vanLeeuwen2007} Hipparcos parallaxes,
\citealt{Groenewegen2008} found that the Hipparcos red clump giants now give
$M_{K_s}\textrm{(RC)}=-1.54\pm0.05$ ($M_{K}\textrm{(RC)}=-1.50\pm0.05$) over
$-0.9\le \mathrm{[Fe/H]}\le0.3$, with a selection bias, whereby accurate K-magnitudes are only available for relatively few bright stars, meaning the actual value is likely brighter. The newer values hold better agreement
with the theoretical results of \citealt{Salaris2002}, who derived an average
value for the solar neighbourhood of $M_{K_s}\textrm{(RC)}=-1.58$, derived
via modelling with a SFR and age-metallicity relation.

Given the disagreement over the $K$-band magnitude for the clump and possible metallicity/age variations, we investigated the use of the three normalizations for $M_{K_s}$ for our derivation of the RAVE red clump distances. Table \ref{tab1} summarizes these values, where the Version A is the standard value from \citealt{Alves2000} and \citealt{Grocholski2002}, while Version B is the new value from \citealt{Groenewegen2008}. Version C is derived from the theoretical models of \citealt{Salaris2002} and attempts to take into account some of the possible systematics for lower metallicity stars, where we use the variation of age and metallicity of RAVE stars described above (i.e., $M_{K_s}(RC)=-1.64$ at $|Z|=0\,\kpc$ to $M_{K_s}(RC)=-1.39$ at $|Z|=4\,\kpc$) to develop a simple linear relation of $M_{K_s}$ with $Z$, where the value for each star is derived iteratively. In Section \ref{sec4.1.2} we examine the effect of the different normalizations on our results. 

\begin{figure*}
\begin{center}
\includegraphics[width=\linewidth]{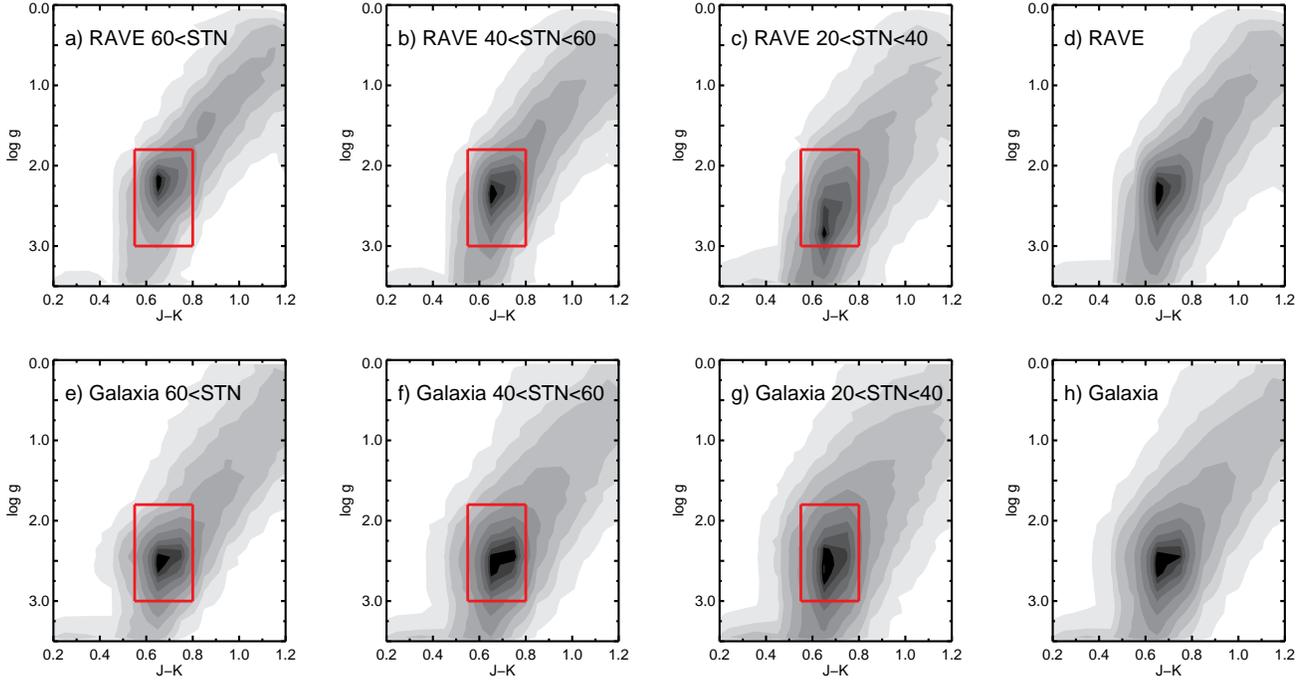}
\caption{The $J-K$-$\log g$ plane for RAVE giants for three STN regimes (a - c) and the corresponding \textsc{Galaxia} models (e-f). Panels (d) and (h) show the aggregate RAVE and \textsc{Galaxia} samples respectively. The red clump is visible as an over-density centred at  $(J-K,\ \log g)=(0.65,\ 2.2)$ and the selection for the red clump region is outlined by the red box. There is a larger movement of the red clump position in $\log g$ with decreasing STN for RAVE stars compared to \textsc{Galaxia}.}
\label{fig1}
\end{center}
\end{figure*}

\subsection{The \textsc{Galaxia} red clump}
\label{sec3.2}

\begin{figure*}
\begin{center}
\includegraphics[width=\linewidth]{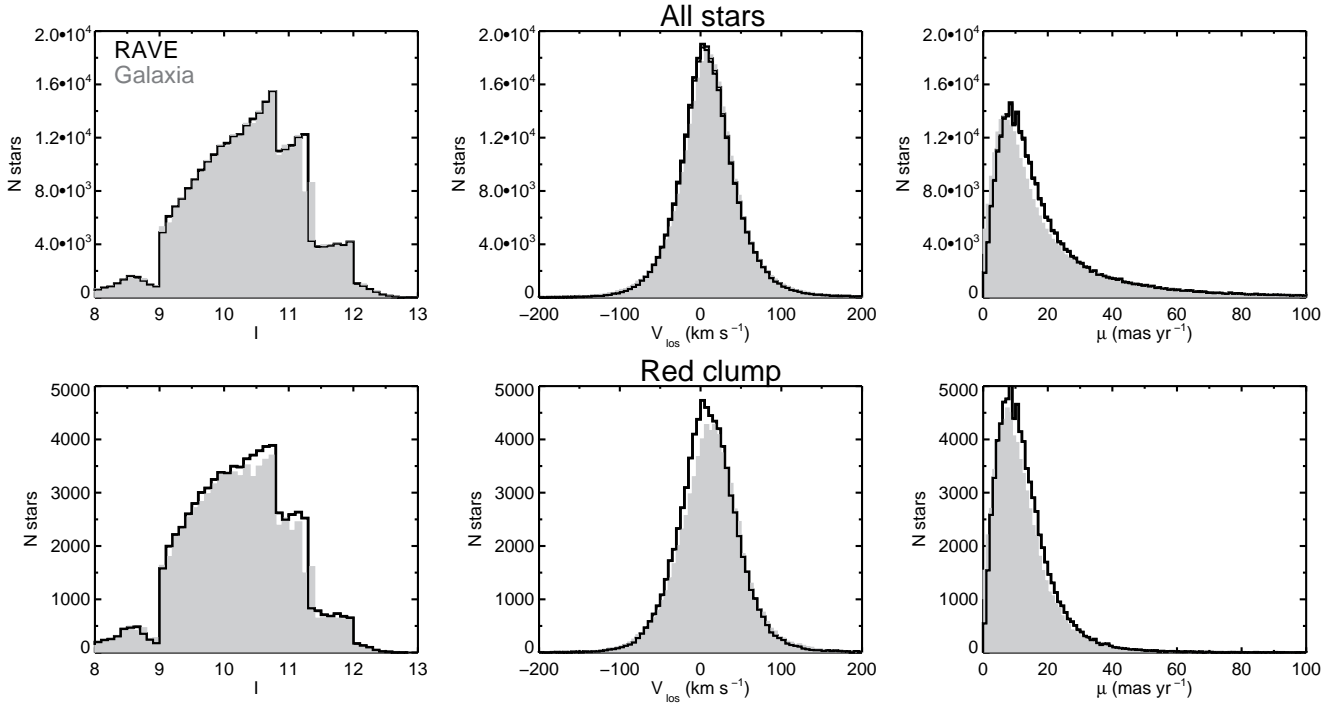}
\caption{The distributions for $I$, $V_\mathrm{los}$ and overall proper motion $\mu$ for the entire sample of unique, cleaned RAVE stars between $8<I<13$ (top) and those selected as part of red clump selected stars (bottom). The $I$ magnitudes for the RAVE stars are from DENIS, apart from those with problematic values as discussed in Section \ref{sec3.2}. The \textsc{Galaxia} model reproduces the two observational kinematic distributions.}
\label{fig2}
\end{center}
\end{figure*}

To establish the best selection method for the red-clump stars we first tried
to match the observed distribution of the selected RC stars with the
prediction of \textsc{Galaxia}. The errors in RAVE stellar parameters decrease
with STN, so the red clump is more localized at higher STN values.  Further,
the errors in $\teff$ and $\log g$ are correlated. To account for this in the
modelling of the clump, we split the data into three regimes: STN values
between $20< \mathrm{STN}<40$, $40< \mathrm{STN}<60$ and $60< \mathrm{STN}$.
We then generated \textsc{Galaxia} models for each STN regime for $5^\circ \times
5^\circ$ squares, matching the I-band distribution of  \textsc{Galaxia} to that of RAVE
in bins of 0.2 mag. The three models for each STN slice of the data were added together to give an overall Galaxia model, with an STN value of 20, 40 or 60 assigned to the \textsc{Galaxia} `stars' to indicate which STN regime they were generated from. Note that, as stated in the DR2 and DR3 releases, the DENIS
I-band photometry has large errors for a significant fraction of the stars.
We therefore used Equation 24 from the DR2 paper to calculate I magnitudes
for those RAVE stars which do not satisfy the condition
$-0.2<(I_\mathrm{DENIS}-J_\mathrm{2MASS})-(J_\mathrm{2MASS}-K_\mathrm{2MASS})<0.6$,
as well as those that have 2MASS photometry but no DENIS values. For the
models the reddening at infinity is matched to that of the value in Schlegel
map and to convert $E(B-V)$ to extinction in different photometric bands we
used the conversion factors in Table 6 of \citealt{Schlegel1998}. 

Errors were then added to the values of $\teff$, $\log g$ and $J-K$ from \textsc{Galaxia} to match the distributions in the RAVE data. The same seed was used
for the random number generator for the temperature and gravity values to
account for the error covariance. Table 2 gives the standard deviations of
the errors that were added to the values from \textsc{Galaxia}. The average
$J-K$ error, $0.01$, is smaller than the expected average observational error
in $J-K$, $\sqrt{eJ^2+eK_S^2}=0.03$. An error of $0.01$ in $J-K$ corresponds
to an error in $\teff$ of $\sim40\,K$ \citep{Alonso1996}. Given this small
value and that, unlike $\teff$, the spread in $J-K$ does not change with STN,
we opted to select the RAVE red clump in $(J-K,\log g)$.

\subsection{Selection of RC stars}
\label{sec3.3}

Figure \ref{fig1} plots the $(J-K,\log g)$ plane for the RAVE giants within
$8<I<13$ in the three STN regimes along with the corresponding \textsc{Galaxia} predictions including errors. Here we see the clump at
$J-K\sim0.65$ and $\log g\sim2.2$. There are some notable differences between
the models and the observed distributions. First, the position of the clump
in $\log g$ decreases with decreasing STN for the RAVE stars. This change is
not matched by \textsc{Galaxia}: the predicted distribution has mean $\log
g\sim2.4$. With increasing dispersion at lower STN it elongates but does not
shift.  Thus, the effect seen in the RAVE data appears not to be astrophysical in origin
but rather a result of the parameter determination from RAVE spectra (for a discussion on systematic trends of stellar parameter estimates with SNR see DR3). Also,
there is an absence of horizontal-branch stars below $J-K<0.55$ in the RAVE
data relative to the prediction by \textsc{Galaxia}. This latter difference does
not affect our results as they are not included in the selection region.

 \begin{table}
\begin{footnotesize}
\begin{tabular}{lllll}
\hline
STN regime&$\sigma(\log g)$&$\sigma(\teff)$&$\sigma(J-K)$\\
 \hline
$60\le \mathrm{STN}$&0.25&25&0.01\\
$40\le \mathrm{STN}<60$&0.35&110&0.01\\
$20\le \mathrm{STN}<40$&0.45&150&0.01\\
\end{tabular}
\caption{Standard deviations of the error distributions added to the \textsc{Galaxia}
 model to match the RAVE distributions in the three STN regimes. The same seed was used for the $\log g$ and $\teff$ to mimic the error covariance.}
\label{tab2}
\end{footnotesize}
\end{table}

We select the red clump as those stars within $0.55\le J-K\le0.8$, $1.8\le \log g\le3.0$. We do not include a STN dependence as extending the selection box further up in gravity will mean an increased fraction of sub-giants in the sample and thus erroneous distances. The red box in Figure \ref{fig1} displays the selection criteria, where we see that for $20<\mathrm{STN}<40$ some of the clump is cutoff at higher values of $\log g$. Applied to our entire cleaned RAVE set, this selection yields 78,019 stars. A sample of red clump stars were also selected from the \textsc{Galaxia} model in the same way, with a slightly smaller number of 73,594 `stars'.

Figure \ref{fig2} gives the distributions of the entire RAVE data set and red
clump stars in $I,\ V_\mathrm{los}$ and overall proper motion,
$\mu=\sqrt{\mu_\alpha^2+\mu_\delta^2}$. The  distributions from \textsc{Galaxia} are also given, with an additional spread of $2\,\kms$ and $2.7\,\masyr$ added to
the \textsc{Galaxia} line-of-sight velocity and proper motion, respectively, to simulate
the observational uncertainties. These values give the average value of the error in these quantities for the RAVE catalog stars. Here we see that \textsc{Galaxia}
reproduces quite well the basic distributions of the two observational
kinematic parameters which affirms our use of it in comparisons. The small difference in the proper motion distribution could suggest that the actual errors in proper motion are slightly larger than estimated.

\section{Error analysis}
\label{sec4}

The most problematic errors are systematic ones. There are multiple ways that these can be introduced into the kinematics. We examine these in some detail before discussing measurement errors in Section \ref{sec4.2}, Poisson noise in Section \ref{sec4.3} and the final data cuts in Section \ref{sec4.4}.

\subsection{Systematic error sources}
\label{sec4.1}

\subsubsection{Contamination by first-ascent giants}
\label{sec4.1.1}

In the $(J-K,\log g)$ plane the red clump is overlaid on the
first-ascent giant branch (see Figure \ref{fig1}). Our selection criteria therefore also select
first-ascent giants, including some sub-giants, which can lead to
distance and velocity errors. To assess how this contamination affects our
distances and kinematics, we use the red-clump selected stars from the \textsc{Galaxia} model and compute a distance to them assuming a single red-clump magnitude.
These results can then be compared to the true values for the model stars.
\textsc{Galaxia} uses mainly Padova isochrones and with a median red-clump K-band
magnitude of $M_{K_s}\textrm{(RC)}=-1.50$. This is slightly different to that
applied above to our data, but is not significant as we are only performing an
internal comparison with \textsc{Galaxia}.

\begin{figure*}
\begin{center}
\includegraphics[width=\linewidth]{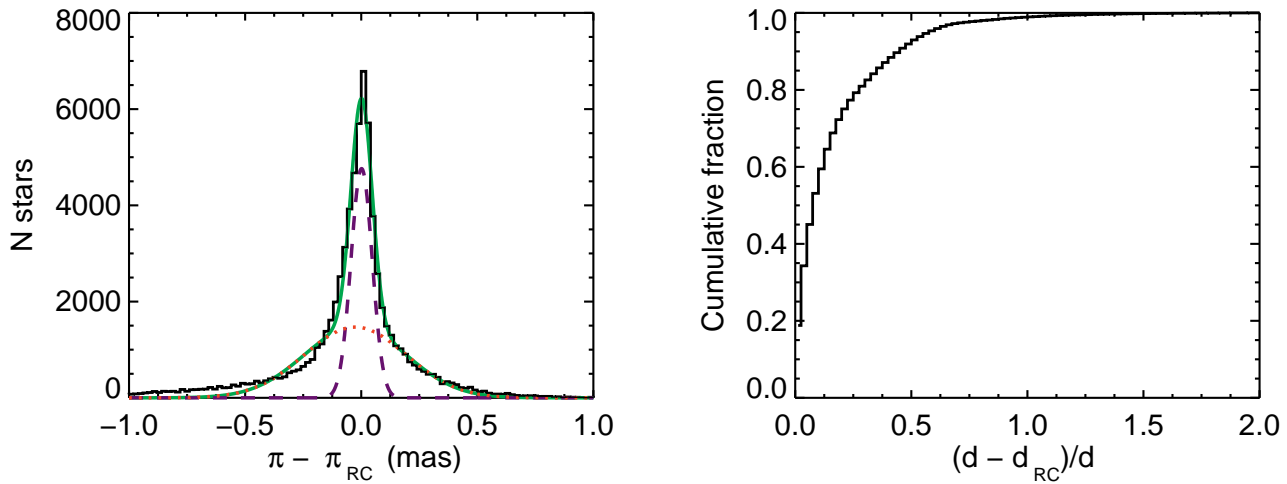}
\caption{a) Histogram of parallax error for \textsc{Galaxia} model stars caused by assuming a single red clump magnitude for stars selected in the red clump region. The distribution is decomposed into two Gaussians: one associated with the true RC (dashed blue line) and one with the first ascent giants (dotted red line). Their sum is given by the solid green line. b) The cumulative distribution of the corresponding distance errors. }
\label{fig3}
\end{center}
\end{figure*}

Figure \ref{fig3}a shows the histogram of the parallax errors. The distributions can be
decomposed into two Gaussians; one representing the red
clump and the other first-ascent giants. Note that the histogram of distance errors does not have such a tidy Gaussian decomposition, hence why we work with the parallax errors. The spread in the
red-clump values is caused by the clump's distribution in age and
metallicity; the average $M_K$ value chosen is not true for all red-clump
stars. The first-ascent giants have larger errors because while
their $\log g$ values may overlap with the clump, their absolute $K$-band
magnitude can be quite different. From this decomposition we estimate that
the $\sim40\:\mathrm{per\:cent}$ of selected stars are actually red-clump stars and the rest
are first-ascent giants. Fortunately however, the mean of the
background distribution is very similar to that of the red clump, with a
systematic shift of only $\sim2\:\mathrm{per\:cent}$ in distance. In Figure
\ref{fig3}b we plot the cumulative distribution of the distance
errors. From this we can see that despite the high level of contamination,
$80\:\mathrm{per\:cent}$ of stars have distances errors of less that $25\:\mathrm{per\:cent}$. 

\begin{figure*}
\begin{center}
\includegraphics[width=\linewidth]{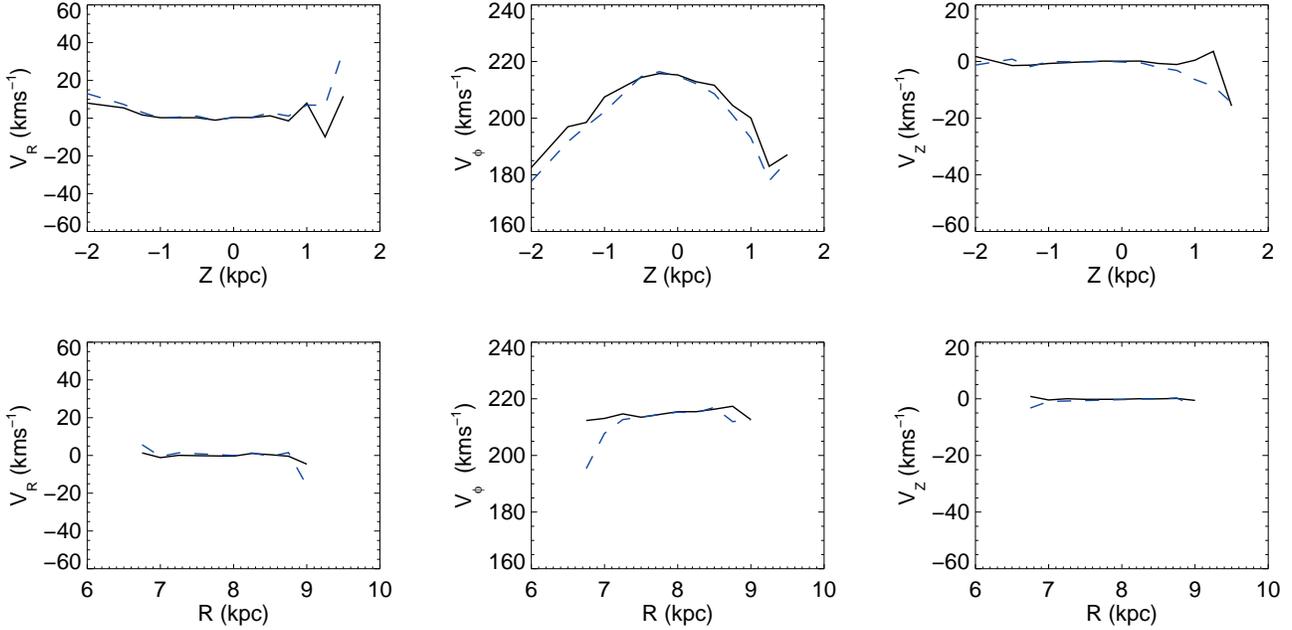}
\caption{Comparison between simulated \textsc{Galaxia} model results using the true \textsc{Galaxia} model distances (solid black line) and those assuming a RC $M_K$ magnitude (dashed blue line). The average $\VR$, $\VPHI$ and $\VZ$ for stars within $7.5<R<8.5\,\kpc$ are plotted as a function of $Z$ (top), as well as the values for $|Z|<0.5\,\kpc$ as a function of $R$ (bottom).} 
\label{fig4}
\end{center}
\end{figure*}

To establish just how these distance errors affect our results, we calculated
$(\VR,\VPHI,\VZ)$ for stars in the solar cylinder ($7.5<R/\kpc<8.5$) from
the pseudo-data produced by \textsc{Galaxia} using both the true distance to
each star, and the distance one infers from the star's apparent magnitude and
a single RC $M_K$.  The upper panels of Figure \ref{fig4} show
the resulting plots of mean velocity components as a function of $|Z|$. In
the panels for $\VR$ and $\VZ$ the
differences between the velocities from true distances (solid line) and from
ones derived from a single absolute magnitude (dashed line) are noticeable
only at the limits of the surveyed region. In the case of 
 $\VPHI$ the single value of $M_K$ leads to a consistent underestimation by
$\sim5\,\kms$.  The lower panels of Figure \ref{fig4} show the
corresponding results for stars that lie within $0.5\,\kpc$ of the plane,
binned in $R$. Again the impact on $\VR$ and $\VZ$ of using a single absolute
magnitude is evident only at the survey limits. We conclude that the use of a
single value of $M_K$ does not significantly compromise our results.

\subsubsection{$M_K$ normalisation}
\label{sec4.1.2}

\begin{figure*}
\begin{center}
\includegraphics[width=\linewidth]{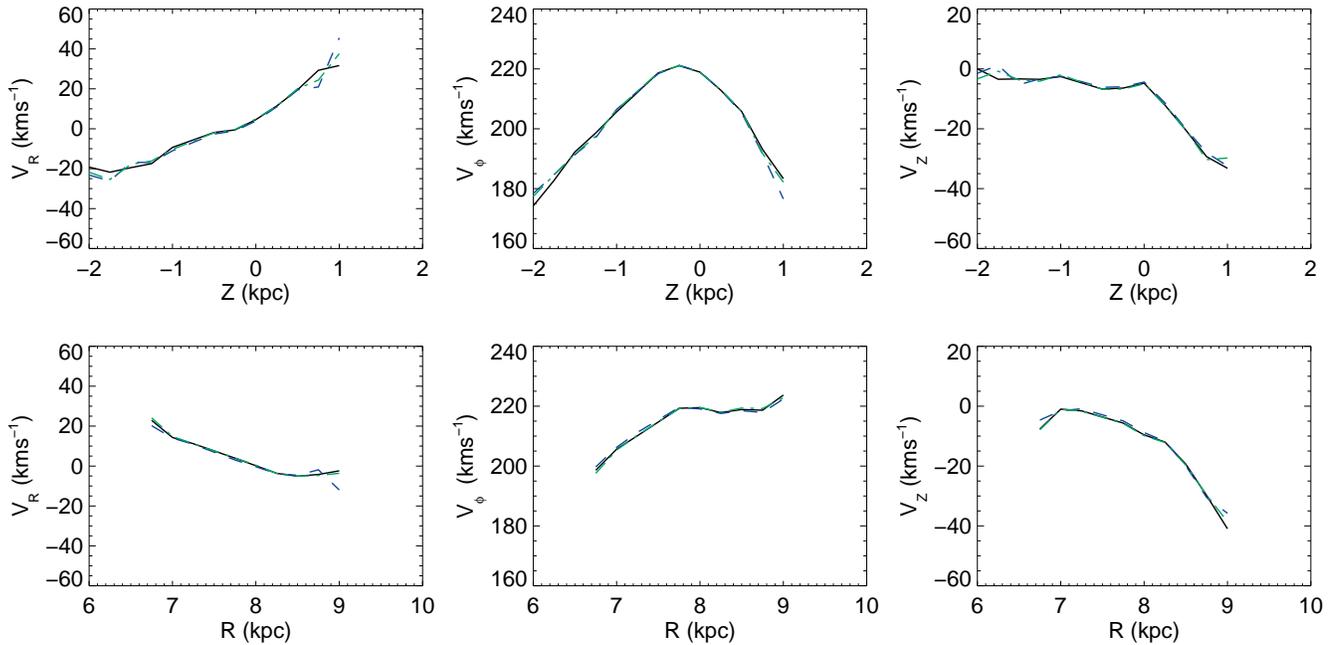}
\caption{Comparison between the three version of the red clump $M_K$ normalization. The average $\VR$, $\VPHI$ and $\VZ$ for RAVE stars within $7.5<R<8.5\,\kpc$ are plotted as a function of $Z$ (top), as well as the values for $|Z|<0.5\,\kpc$ as a function of $R$ (bottom). Normalization A from Table \ref{tab1} is given by the solid black line, normalization B by the blue dashed line and normalization C by the green dot-dashed line. The results are relatively insensitive to the assumption of a single $M_K$ magnitude.} 
\label{fig5}
\end{center}
\end{figure*}

In Section \ref{sec3.1} we introduced three different normalizations for the
$M_K$ value for the red clump. Figure \ref{fig5} shows the effect of
using these different normalizations on the average values of $\VR,\ \VPHI,\
\VZ$ for stars in the solar cylinder (top row of panels) and for stars within
$0.5\,\kpc$ of the plane (lower row).  We see that the differences in
distance produced by the three normalizations has an insignificant impact on
the average velocities: only at the limits of the survey, where sample sizes
are small, do the differences
reach $15\,\kms$.  Thus the dispersion in the
velocity errors, on the order of $10\,\kms$, caused by the single $M_K$ assumption are averaged over and thus
all but vanish. It is only those bins - the farthest ones - with the lower number of
stars that show any effect. 

Given that the effect of the absolute-magnitude normalisation is minor, we use
only normalization A for the remainder of this analysis.

\subsubsection{Proper motions}
\label{sec4.1.5}

RAVE proper motions are derived from several sources, although the majority
of the catalog gives PPMX \citep{Roeser2008} or UCAC2 \citep{Zacharias2004}
proper motions. To investigate the contribution of systematic proper-motion
errors, we cross-matched with two additional proper-motion catalogs, the
third US Naval Observatory CCD Astrograph Catalog, UCAC3
\citep{Zacharias2010}, and the Yale/San Juan Southern Proper Motion Catalog,
SPM4 \citep{Girard2011}. The UCAC3 proper motions suffer from strong
plate-dependent systematic distortions north of $\delta=-20^\circ$
\citep{Roeser2010}, and so we exclude stars with $\delta>-20^\circ $ in this
catalog.

\begin{figure*}
\begin{center}
\includegraphics[width=\linewidth]{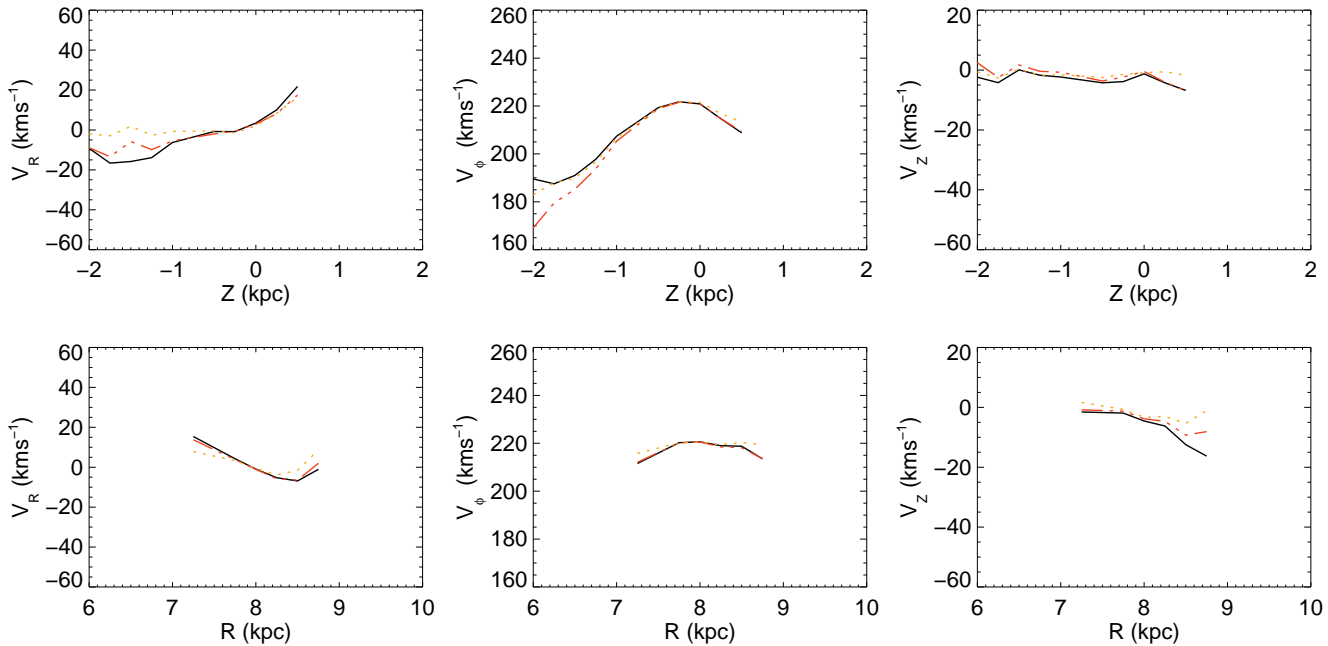}
\caption{As in Figure \ref{fig5}, but a comparison between the results using different proper motion sources for RAVE stars with $\delta<-20^\circ,\ -1<Y<0\,\kpc$, with RAVE catalog proper motions (black solid line), UCAC3 proper motions (dotted orange line) and SPM4 proper motions (dash-dottted red line).}
\label{fig6}
\end{center}
\end{figure*}

Figure \ref{fig6} shows the mean values of $\VR$, $\VPHI$ and $\VZ$ with
$R$ and $Z$ for stars below $\delta=-20^\circ,\ -1<Y<0\,\kpc$ calculated
using the three sources of proper motions, where the restrictions in $\delta$
and $Y$ were made to keep as uniform as possible the fractions of stars that
have entries in each of the three catalogs.
The mean velocities do change with proper-motion source, with the
discrepancies
largest between RAVE proper motions and those from the SPM4 catalog. The
divergence is largest for $Z<-1\,\kpc$ for $\VR$ and $\VPHI$ with differences
up to $20\,\kms$ in $\VPHI$. Some of these variations may be the result of
different coverage of the sample volume -- which we have tried to minimize
with the restrictions on $\delta,\ Y$ above. Nevertheless, it does indicate
that the proper motions are a significant source of systematic error. 

It is difficult to say \textit{a priori} which catalog is closer to the
truth; a detailed comparison is beyond the scope of this paper. Given that
the systematic differences between proper-motion catalogs can cause
significant differences in the derived velocities, we use all three proper
motion catalogs in our further analysis, concentrating on the two most
divergent proper motions, namely those in the RAVE and SPM4 catalogs.

\subsubsection{Binary stars}
\label{sec4.1.3}

Two types of spectroscopic binaries can potentially affect our results;
single-lined spectroscopic binaries (SB1) stars and double-lined
spectroscopic binaries (SB2) stars. Both kinds introduced an additional
velocity variation to the sample, with the magnitude much larger for SB2
stars. For the SB2 stars, there is the additional problem that the RAVE
processing pipeline assumes that each observed spectrum is that of a single
star.

\citealt{Matijevic2011} uses repeat observations to estimate the fraction of
single-line spectroscopic (SB1) binary stars in the RAVE sample, finding a
lower limit of $10-15\:\%$ of the sample consists of SB1 stars. The technique is
biased towards shorter period binaries, but can be used to gauge the
contribution to the measured 
dispersions from SB1 stars. In \citealt{Matijevic2011} the
mode of the distribution in the velocity variation from the SB1 stars is
$6\,\kms$. Undetected, long period binary stars will have smaller velocity variations. Thus, the small additional velocity dispersion on a limited fraction
of the red clump stars from SB1 sources can be neglected, especially as we are measuring mean
velocities in this paper. The fraction of SB2 stars in RAVE is much smaller;
\citealt{Matijevic2010} estimates $0.5\:\%$ of the sample are SB2 stars. Many of
these are flagged however, as well as any other peculiar stars, and have been
removed from the sample. Unresolved SB2 stars are fortunately rare
\citep{Matijevic2010} and these and the remaining SB2 stars do not affect our
results substantially.

\subsubsection{Extinction correction}
\label{sec4.1.4}

As in \citealt{Williams2011}, extinction is calculated iteratively from the distances using \citealt{Schlegel1998} dust maps and assuming a Galactic dust distribution as in \citealt{Beers2000}. Note that the maps were adjusted using the correction of \citealt{Yasuda2007}. The extinction reaches relatively large values of $A_K>0.1$ for $|b|<10^\circ$. However, the effect on the velocities is minor; the effect of the extinction correction on the total space velocity is greater than $5\,\kms$ in only $1.6\:\mathrm{per\:cent}$ of red clump stars. These stars were excluded from further analysis, leaving 72,365 stars in our red clump sample.

\subsection{Measurement errors}
\label{sec4.2}

The non-systematic sources of errors in the kinematics can be broadly broken into (i) the contribution of random uncertainties in the measurements and (ii) the finite sample size for each volume bin. We will examine each of these in turn, before discussing any kinematic cuts that were applied to the data. 

The Monte Carlo method of error propagation enables an easy tracking of error
covariances, so we employ it here by generating a distribution of 100 test
particles around each input value of distance, proper motion, line-of-sight
velocity\footnote{The positions of the stars are assumed to have negligible
errors.} and then calculating the resulting points in $(\VR,\,\VPHI,\,\VZ)$.
We assume that the proper motions and line-of-sight velocities have Gaussian
errors with standard deviations given by the formal errors for each star. For
the distribution in distance for each star however, we generate a
double-Gaussian distribution in the parallax as in Figure \ref{fig3},
which we then invert to derive the distance distribution. This takes into
account the fact that errors in distance are due both to the intrinsic width
of the RC and the mis-classification of other giants. The ratio between the RC
and first-ascent giants for \textsc{Galaxia} model stars changes with distance, where we found a smaller amount of contaminants for $d>1\,\kpc$. The double-Gaussian in parallax then has dispersions
$(0.05,0.26)\,\mathrm{mas}$ at a ratio of $2:1$ for $d\le1\,\kpc$ and $4:1$
for $d>1\,\kpc$.

\subsection{Poisson noise}
\label{sec4.3}

In each of the bins that are used to calculate the mean velocities, there are
a finite number of stars $N$, so Poisson noise contributes to the errors in
the derived kinematic properties. These errors scale as $1/\sqrt{N}$. To
establish the error contribution from this source we used Bootstrap case
resampling with replacement: for each kinematic quantity we derived a
distribution in the values by randomly resampling from the distribution of
values in the bin. The variance in each value could then be calculated from
the resulting distribution. Poisson errors can dominate over measurement
errors in bins at large distances that contain very few stars.

To ensure that Poisson errors do not dominate our plots, we only use bins which have $N_\textit{stars}> 50$ and only include those points that have errors in the mean of less than $5 \kms$ (see Section \ref{sec4.4}). 

\subsection{Data cuts}
\label{sec4.4}

There are several ways in which to prune the data to those that are deemed
more reliable. However, pruning is liable to introduce kinematic biases. We
investigated the effects on the measured velocities of trimming data via a)
proper motion, b) error in proper motion, c) distance errors, d) magnitude of
total velocity and e) total velocity error, $eV_\mathrm{total}$. In general, we
found that as we increase the cut-off point there is a steady increase in
velocity dispersion and fluctuations in the average velocity, asymptotically
approaching a value as the number of stars approaches the full sample. It is
therefore difficult to justify cuts that remove a significant proportion of
stars. We therefore introduced cuts that remove only the outlier values, as
given in Table \ref{tab3}, and we do not perform a cut on $eV_\mathrm{total}$.

\begin{table*}
\begin{footnotesize}
\begin{tabular}{ll}
\hline
Cut&Reason\\
 \hline
 $\sqrt(\VR^2+(\VPHI-220)^2+\VZ^2) < 600\kms$&Remove outlier velocities\\
$|\VPHI-220| < 600 \kms$&Remove outlier velocities\\
$e_{\mua},\ e_{\mud}< 20\ \masyr$&Remove outlier proper motions\\
$ \mua,\ \mud < 400\ \masyr$&Remove high proper motion stars\\
$e_d/d < 1$&Zwitter only, remove large distance error stars\\
\end{tabular}
\caption{Cuts on the data adopted for this data analysis}
\label{tab3}
\end{footnotesize}
\end{table*}

Since we seek only to follow trends with $R,\,Z$, we do not distinguish
between halo, thick and thin disk stars. The cuts in total velocity therefore
aim to be inclusive of halo stars; the $600\,\kms$ limit is $\sim3\sigma$ the halo dispersion of $213\,\kms$ \citep{Vallenari2006}. An
additional cut in $\VPHI$ is also introduced to limit ourselves to stars that
have plausible rotation velocities.

In our analysis we bin the data in physical space, calculating the
means and dispersion in each bin. A further data cut was performed
post-binning. For each bin we calculate the mean $\langle\VR\rangle,\
\langle\VPHI\rangle,\ \langle\VZ\rangle$.  For each of these we also
calculate an error in the mean, given by standard error propagation as
 \begin{equation}
e_{\langle V_q\rangle}=  {1\over N}\sqrt{ \sum e_{V_q, i}^2},
\end{equation}
 where $q=R,\ \phi,\ Z$ and $i=1,.., N$, with $N$ the number of stars in the
average. The values $e_{V_q, i}$ are given by the MC propagation described in
Section \ref{sec4.2}.  We remove bins with large errors, i.e., $e_{\langle V_q
\rangle} > 5 \kms$. This affects only peripheral points at large
distances from the Sun.

\section{Spatial distribution}
\label{sec5}

\begin{figure*}
\begin{center}
\includegraphics[width=\linewidth]{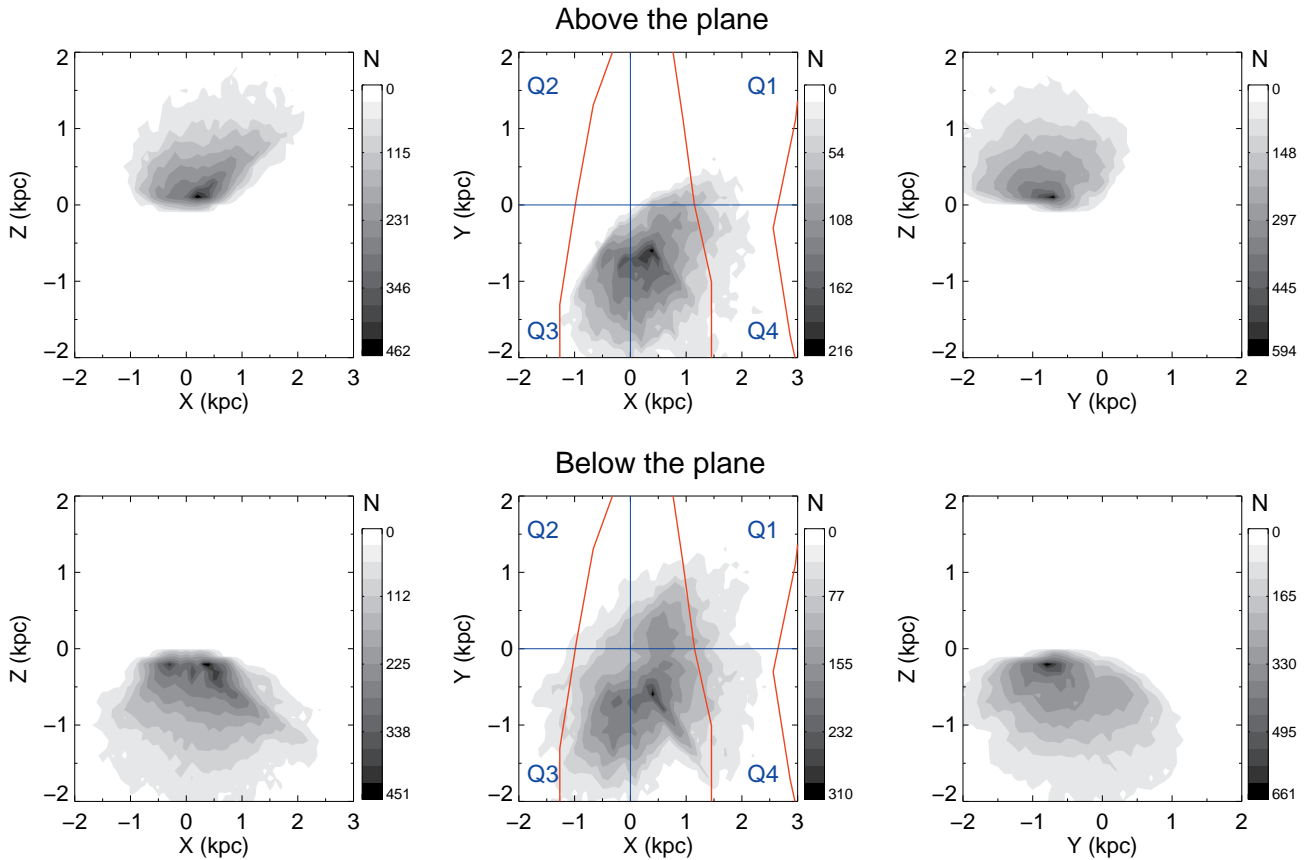}
\caption{The XYZ distribution of RAVE red clump giants, split into samples above the plane (top) and below the plane (bottom). The position of the spiral arms inferred from the CO maps of \citealt{Englmaier2008} are plotted (red lines), giving from left to right the Perseus, Sagittarius-Carina and Scutum-Centurus arms. The four Galactic quadrants are also delimited and labelled (Q1..Q4) in the central XY plot. Note that the diagonal over-density of stars in Quadrant 4 is purely caused by a greater number of observations in that region.}
\label{fig7}
\end{center}
\end{figure*}

Before examining the velocity trends it is helpful to examine the spatial distribution, to see which regions the RAVE red clump selection function samples. Figure \ref{fig7} gives the $XYZ$ distribution for the RAVE red clump stars, where we have differentiated between stars above and below the plane for clarity. Due to RAVE's magnitude limit, the red clump stars sample a region between $0.3<d<2.8\,\kpc$. This, combined with RAVE's uneven sky coverage between the Northern and Southern Galactic hemispheres, means that the sample region is mostly outside $d>0.5\,\kpc$ and Quadrants 1 and 2 above the plane are not sampled.

As in S11, S12, we also plot the location of the spiral arms as inferred from the CO maps of \citealt{Englmaier2008}. Going from outer to inner (left to right), the arms are the Perseus, Sagittarius-Carina and Scutum-Centurus arms. We see that our sample misses the Scutum-Centurus arms entirely, and samples above and below the other two spiral arms.

Another significant nearby feature is the Hercules thick-disk cloud (HTDC)
(\citealt{Larsen1996}, \citealt{Parker2003}. \citealt{Parker2003} detected it via
star counts in the region $l=\pm(20^\circ-55)^\circ$ both above and below
the plane at latitudes $b=\pm(25^\circ-45)^\circ$. This is in
Quadrant 1. Recently,  \citealt{Larsen2011} reported that the HTDC
starts at $(X,\,Y)=(0.5,\,0.5)\,\kpc$ in our co-ordinates, for $0.5<|Z|<1.0\,\kpc$. From Figure
\ref{fig7} we can see that the RAVE sample in the south intersects with
this location of the Hercules thick-disk cloud. \citealt{Juric2008} also found
the HTDC in the location $(X',\,Y',\,Z')=(6.25,\,-2-0,\,1-2)\,\kpc$, corresponding to $(X,\,Y,\,Z)=(1.75,\,0-2,\,1-2)\,\kpc$ in our co-ordiantes. This region is not covered by the
RAVE survey so we cannot see the northern component of the HTDC. In the same
paper, another stellar over-density was found at $(X',\,Y',\,Z')=(9.5,\,0.5,\,1-2)\,\kpc$ ($(X,\,Y,\,Z)=(-1.5,\,-0.5,\,1-2)\,\kpc$), which is just missed by the RAVE volume.

\section{Velocity trends}
\label{sec6}

\begin{figure*}
\begin{center}
\includegraphics[width=1\linewidth]{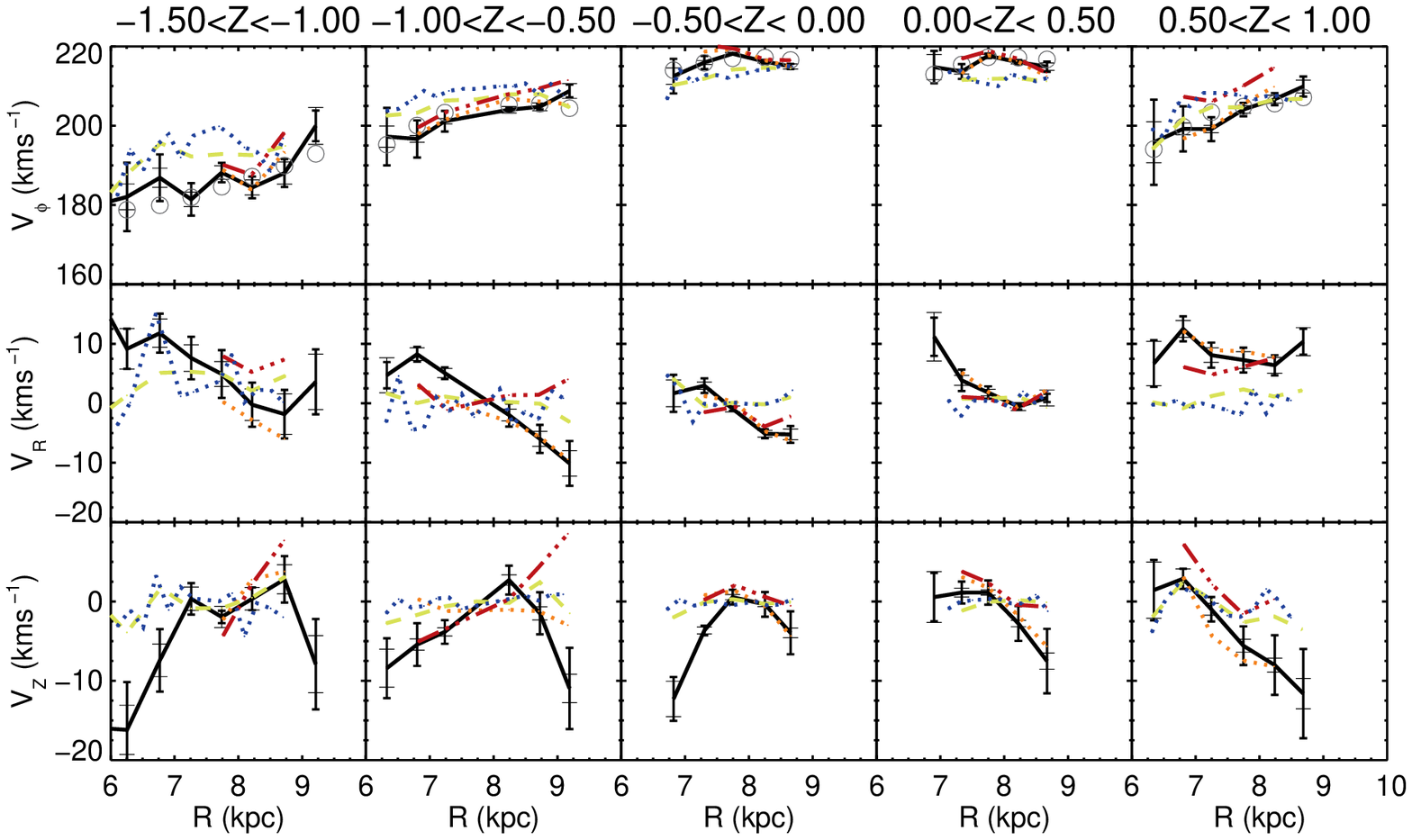}
\caption{The trends in average velocity as a function of ($R,\ Z$) position in the Galaxy for RC stars with RAVE catalog proper motions (black solid line), UCAC3 proper motions (orange dashed line) and SPM4 proper motions (red dot-dashed line). \textsc{Galaxia} model results are also given, using the real \textsc{Galaxia} distances (dark blue) and RC distances (light green). Error bars give the measurement (thick line, short hat) and Poisson (thin line, long hat) errors. The grey open circles in the $\VPHI$ plot give the result of the fit using RAVE catalog proper motions in Section \ref{sec6.2}.}
\label{fig8}
\end{center}
\end{figure*}

Figure \ref{fig8} displays the trends in mean $\VR,\ \VPHI,\ \VZ$ (hereafter we drop the bracket notation for averages) as
functions of $R$ for $0.5\,\kpc$ thick slices in $Z$ using $0.5\,\kpc$ bins
in $R$. Results are shown for three choices of proper motions: those in the
RAVE, UCAC3 and SPM4 catalogs. Also shown
are the results of the pseudo-data from \textsc{Galaxia}: the blue line is obtained using the true distances to pseudo-stars,
while the green line uses distances inferred from RC magnitudes. 
Figures \ref{fig10}, \ref{fig12} and \ref{fig13} display essentially the same results
as contour plots in the $(R,Z)$ plane.  However, to save space we show
results only for the proper motions in the RAVE and SPM4 catalogs and for the \textsc{Galaxia} pseudo-data. The
plotted data are box-car averages over $200\,\pc\times200\,\pc$ wide boxes in
$(R,\ Z)$ with $100\,\pc$ increments in the co-ordinates of the box's centre.  
Finally, Figure \ref{fig14} shows the results
obtained with the proper motions in the RAVE catalog in full 3D -- $\VR,\
\VPHI,\ \VZ$ are averaged over boxes of size $500\times500\times500\,\pc$.
The centres are moved by $250\,\pc$ in $X$ and $Y$ and spaced by $500\,\pc$ in
$Z$.

We now discuss the trends in each velocity component.

\begin{figure*}
\begin{center}
\includegraphics[width=1\linewidth]{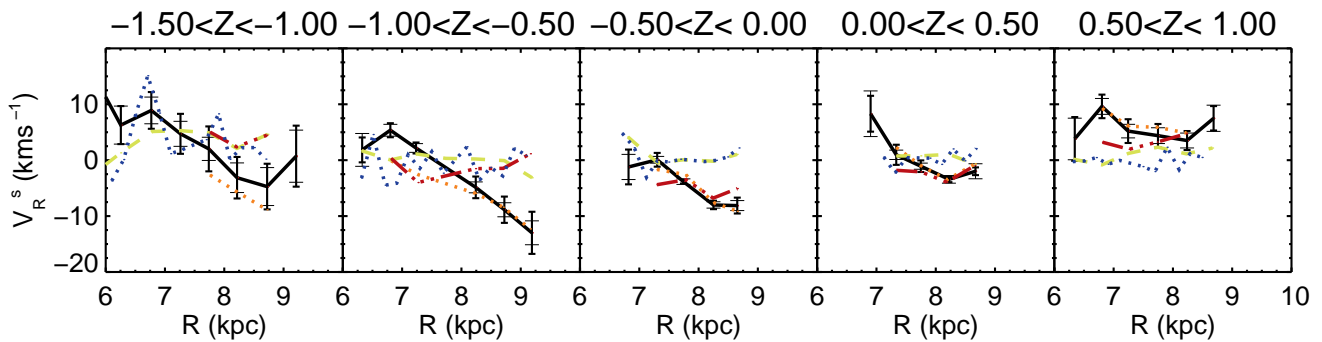}
\caption{As in Figure \ref{fig8} with however $\VR^s$ giving the Galactocentric radial velocity with $U_\odot=14\,\kms$. The values are shifted down with the overall trends unaffected.}
\label{fig9}
\end{center}
\end{figure*}

\begin{figure*}
\begin{minipage}{6cm}
\includegraphics[width=8.8cm]{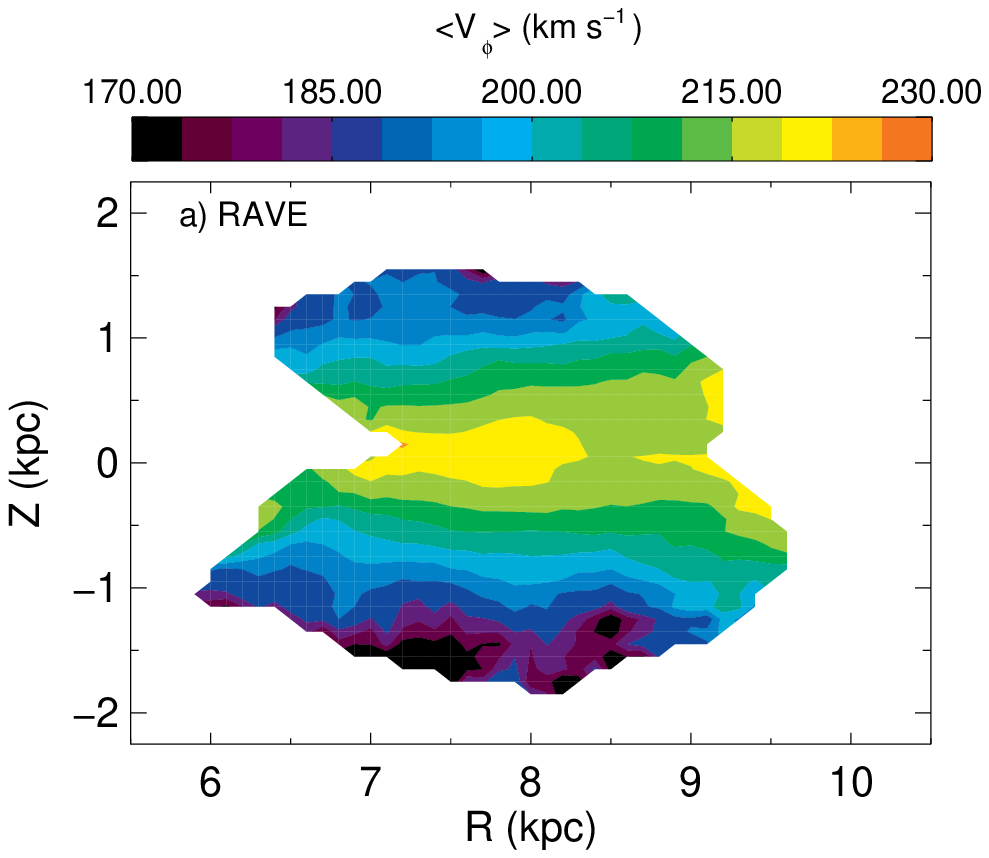}
\end{minipage}
\hfill
\begin{minipage}{9cm}
\includegraphics[width=8.8cm]{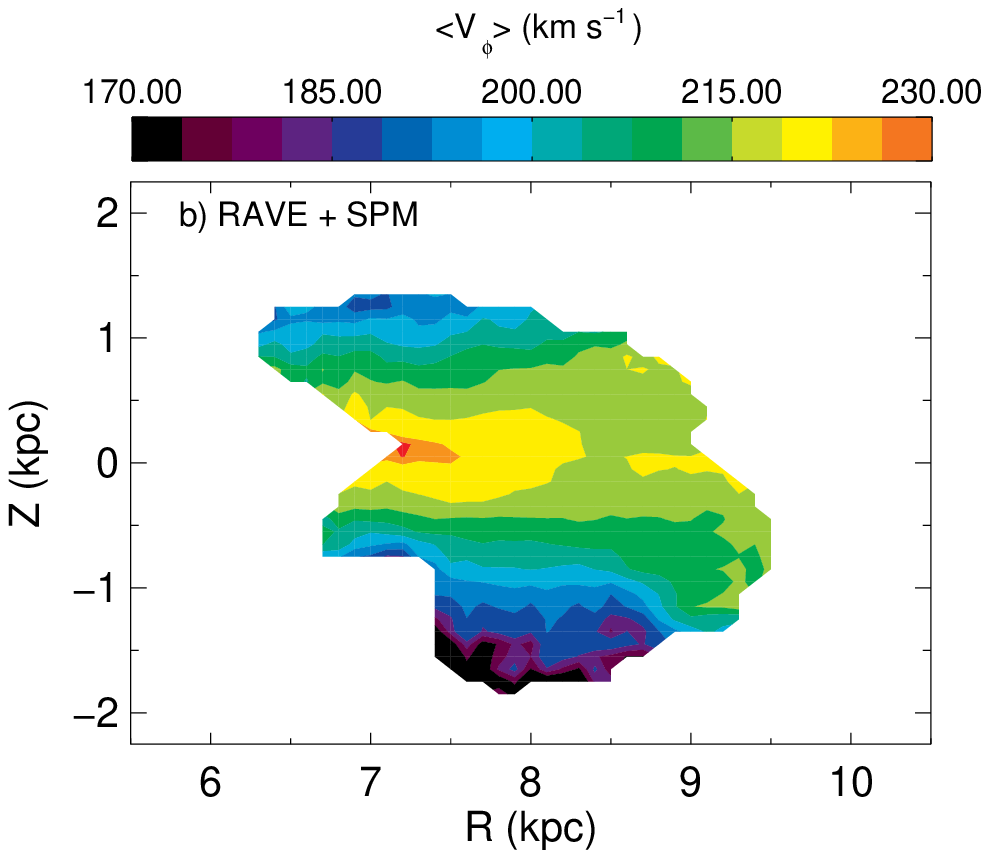}
\end{minipage}
\begin{minipage}{6cm}
\includegraphics[width=8.8cm]{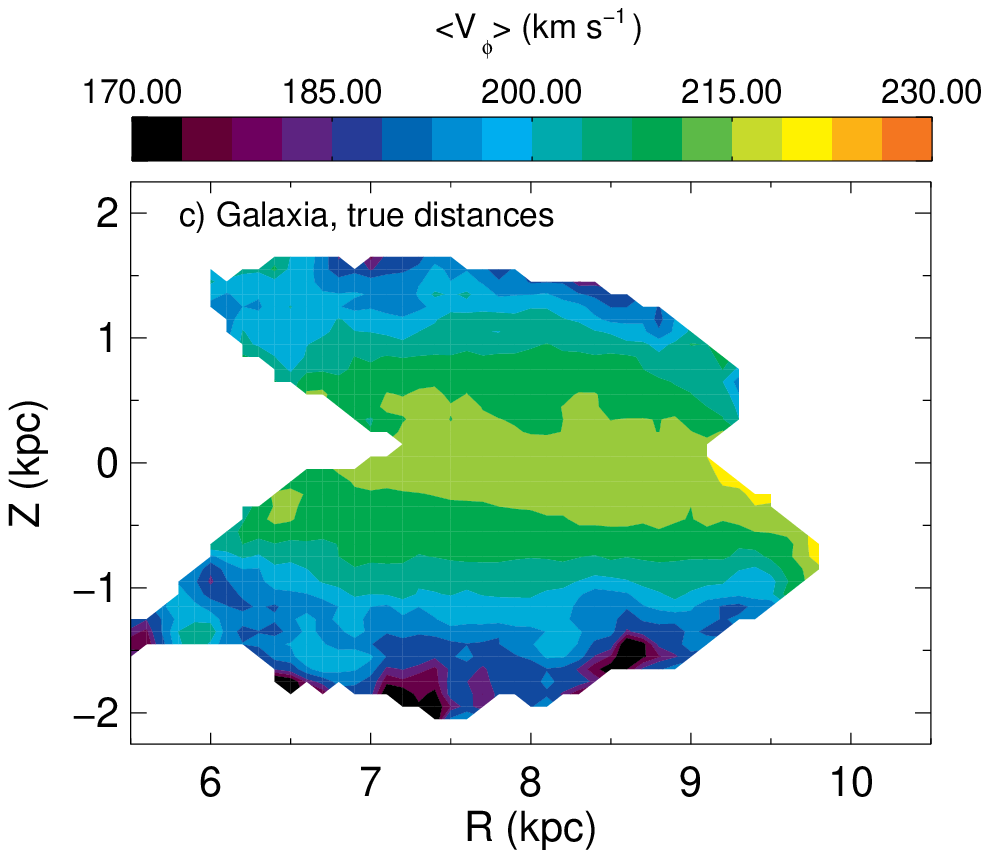}
\end{minipage}
\hfill
\begin{minipage}{9cm}
\includegraphics[width=8.8cm]{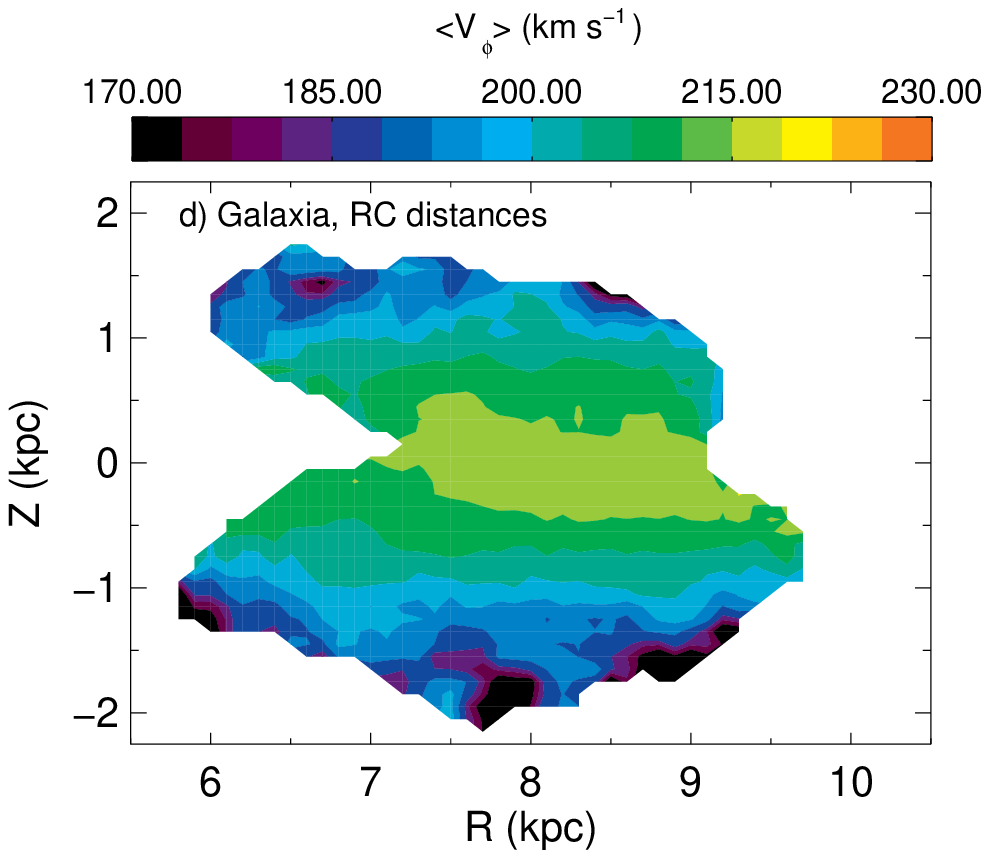}
\end{minipage}
\caption{The trends in average $\VPHI$ as functions of position in the ($R,\ Z$) plane for RAVE RC stars with
RAVE catalog proper motions (a) and SPM4 proper motions (b). \textsc{Galaxia} model results using the ``true" \textsc{Galaxia} distances (c) and the RC-distances (d) are also presented. The
plotted data are box-car averages over $200\,\pc\times200\,\pc$ wide boxes in
$(R,\ Z)$ with $100\,\pc$ increments in the co-ordinates of the box's centre.}
\label{fig10}
\end{figure*}

\begin{figure*}
\begin{minipage}{6cm}
\includegraphics[width=8.8cm]{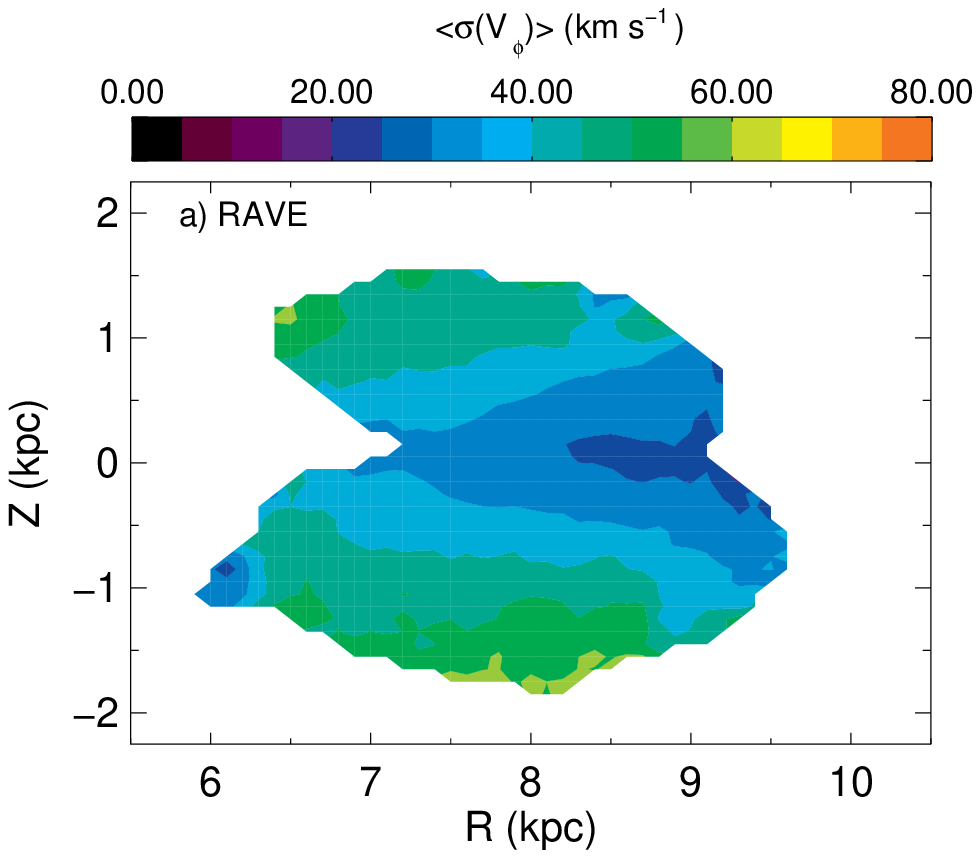}
\end{minipage}
\hfill
\begin{minipage}{9cm}
\includegraphics[width=8.8cm]{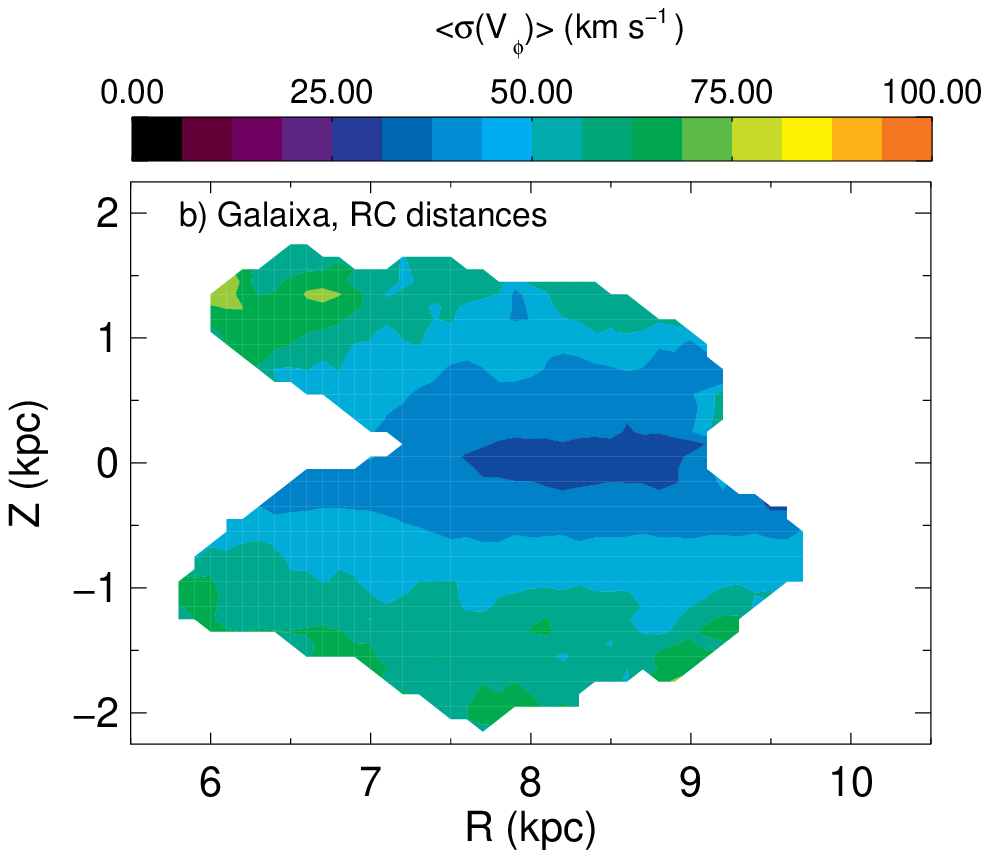}
\end{minipage}
\caption{The trends in average $\sigma(\VPHI)$ as functions of position in the ($R,\ Z$) plane for RAVE RC stars with
RAVE catalog proper motions (a) and for \textsc{Galaxia} model results using the RC distance method (b). The velocity dispersion increases as $\VPHI$ decreases. }
\label{fig11}
\end{figure*}

\begin{figure*}
\begin{minipage}{6cm}
\includegraphics[width=8.8cm]{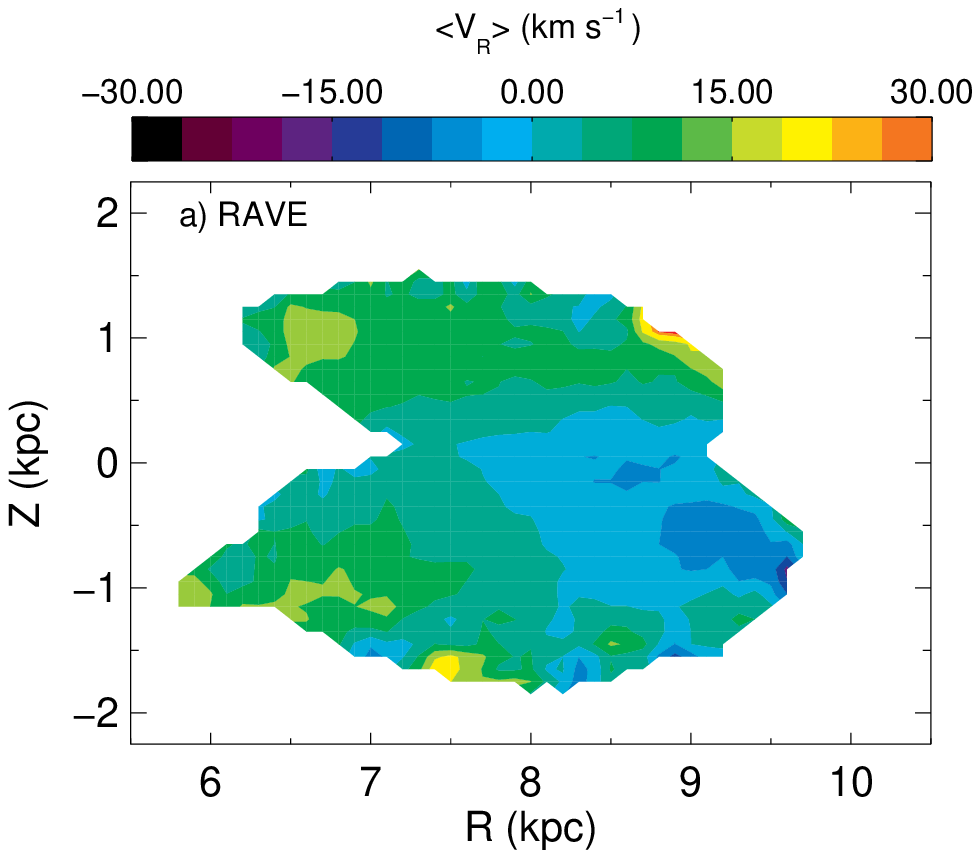}
\end{minipage}
\hfill
\begin{minipage}{9cm}
\includegraphics[width=8.8cm]{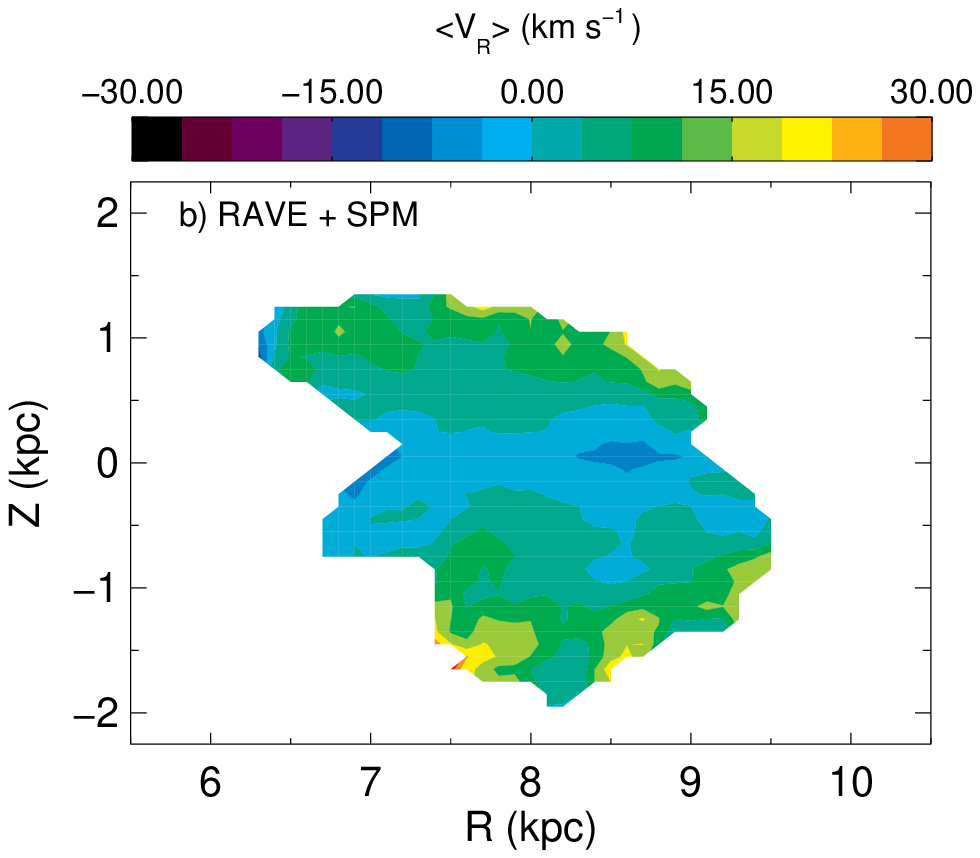}
\end{minipage}
\begin{minipage}{6cm}
\includegraphics[width=8.8cm]{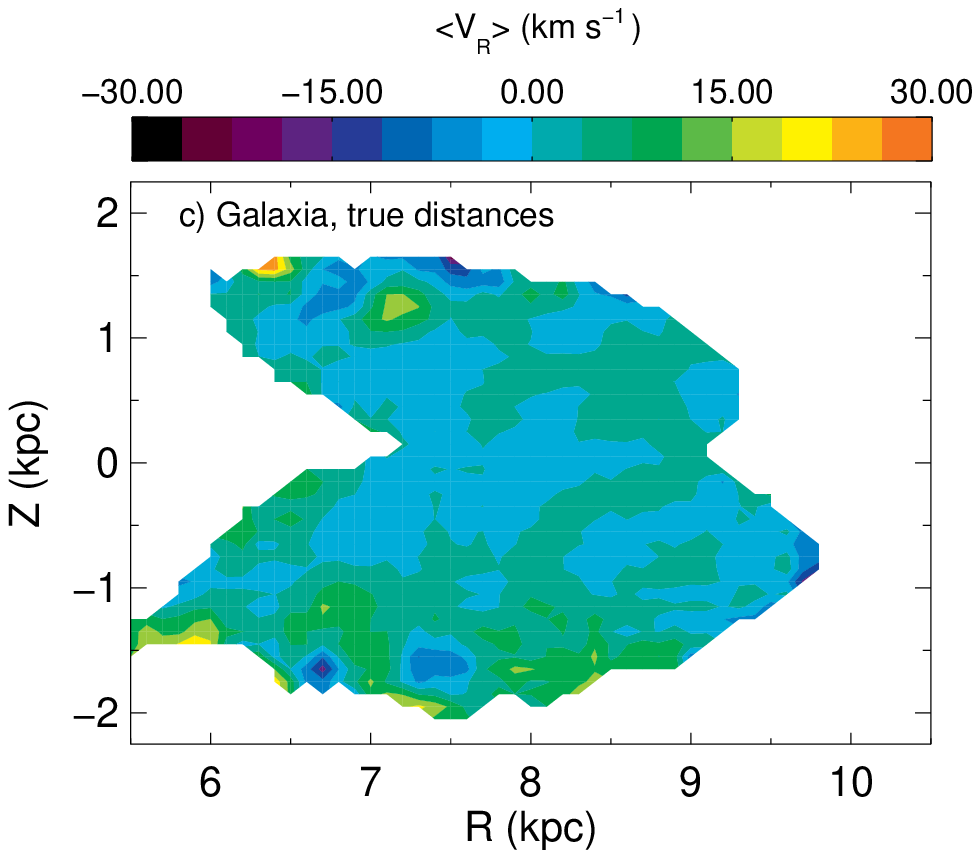}
\end{minipage}
\hfill
\begin{minipage}{9cm}
\includegraphics[width=8.8cm]{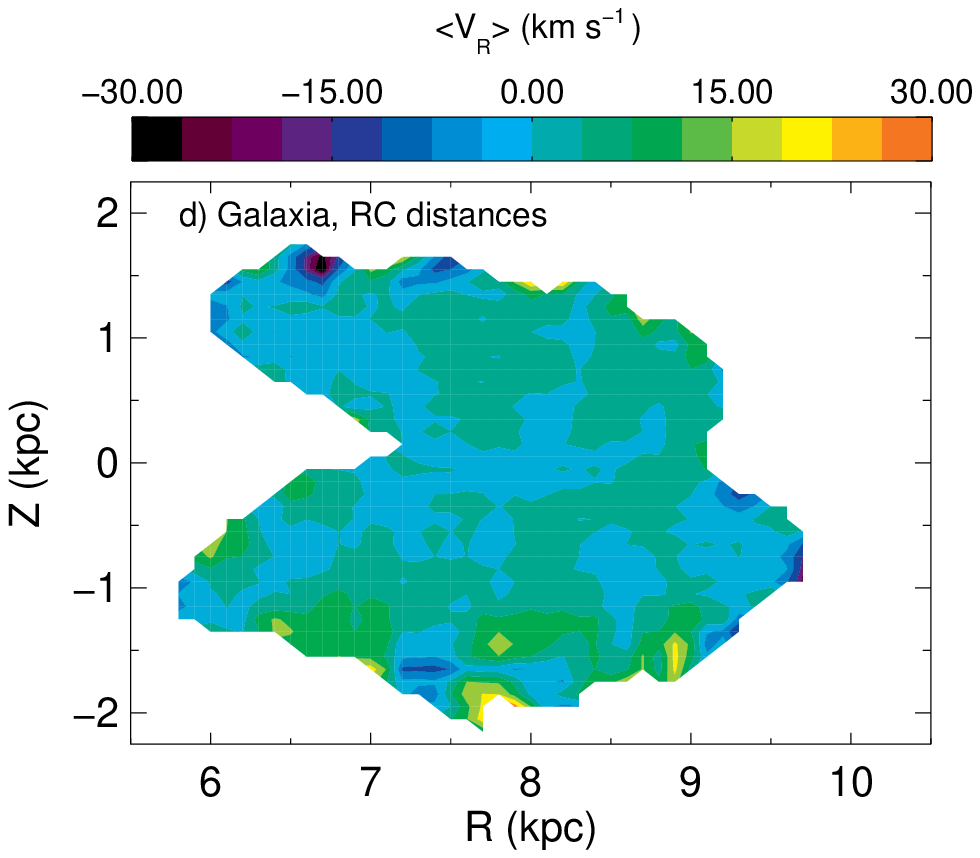}
\end{minipage}
\caption{As in Figure \ref{fig10}, but for $\VR$.}
\label{fig12}
\end{figure*}

\begin{figure*}
\begin{minipage}{6cm}
\includegraphics[width=8.8cm]{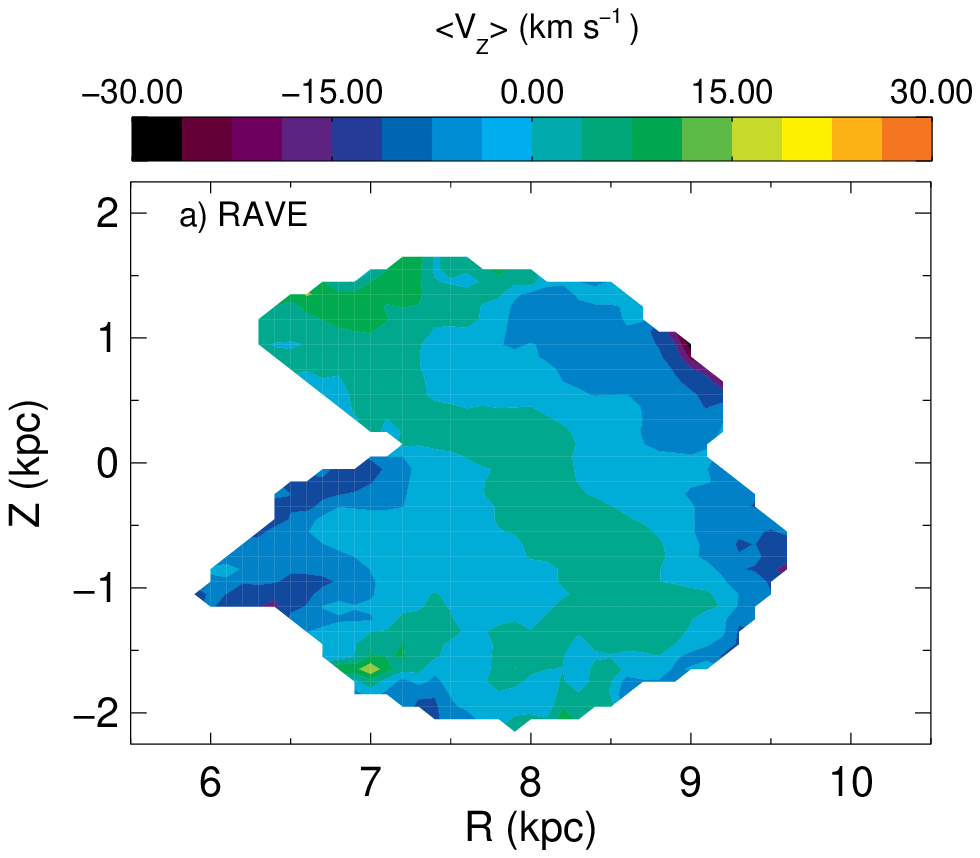}
\end{minipage}
\hfill
\begin{minipage}{9cm}
\includegraphics[width=8.8cm]{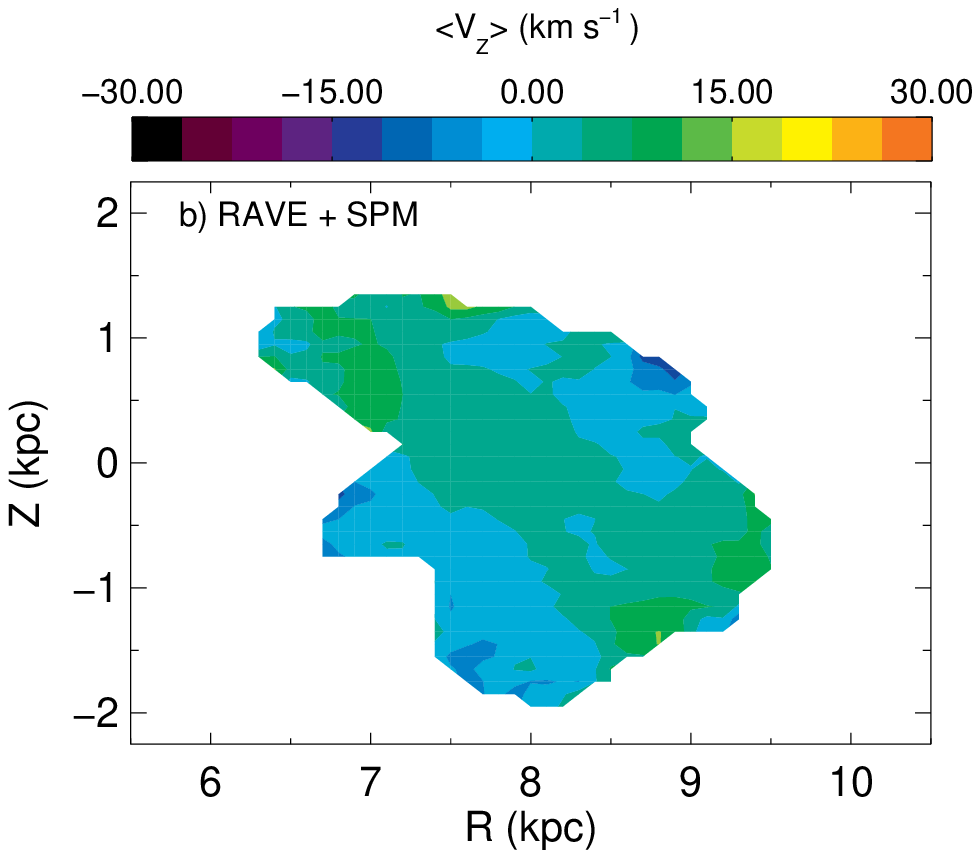}
\end{minipage}
\begin{minipage}{6cm}
\includegraphics[width=8.8cm]{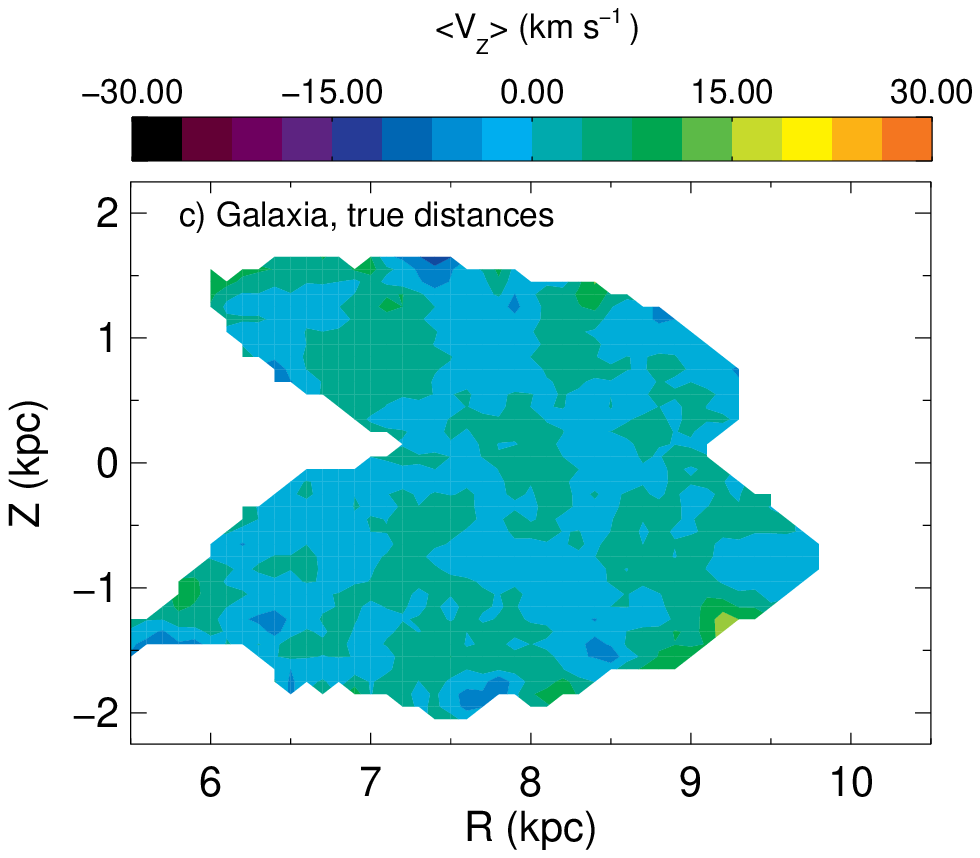}
\end{minipage}
\hfill
\begin{minipage}{9cm}
\includegraphics[width=8.8cm]{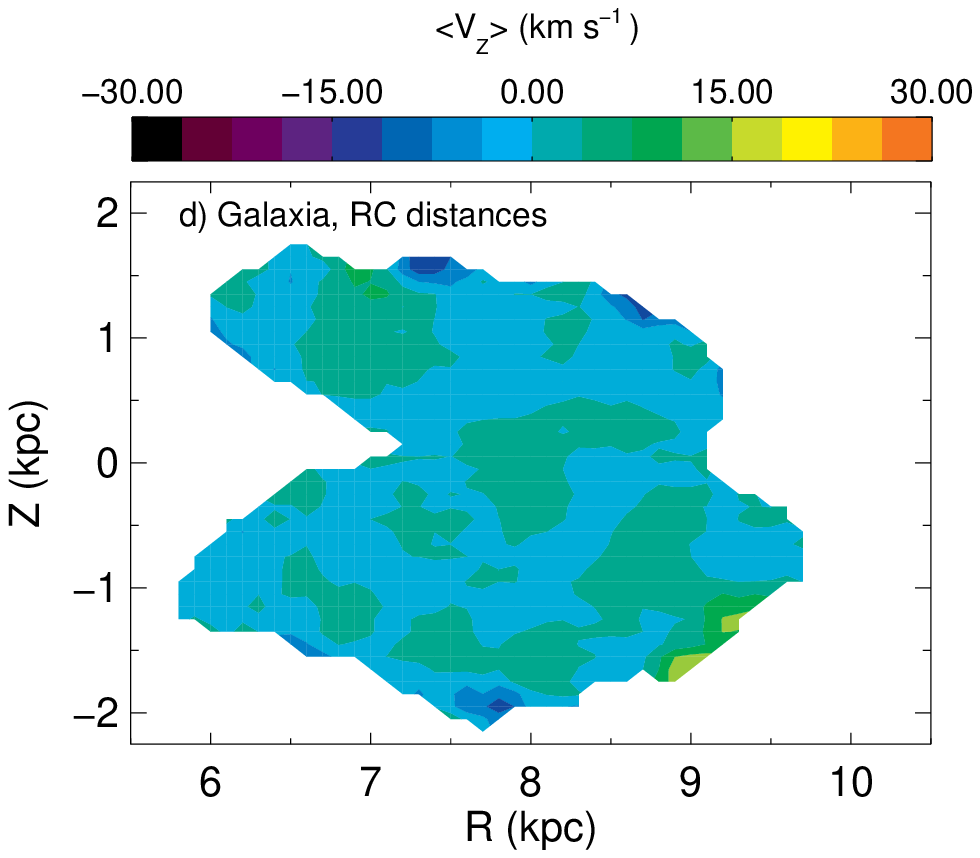}
\end{minipage}
\caption{As in Figure \ref{fig10}, but for $\VZ$.}
\label{fig13}
\end{figure*}

\begin{figure*}
\centering
\begin{tabular}{c}
\includegraphics[width=17cm]{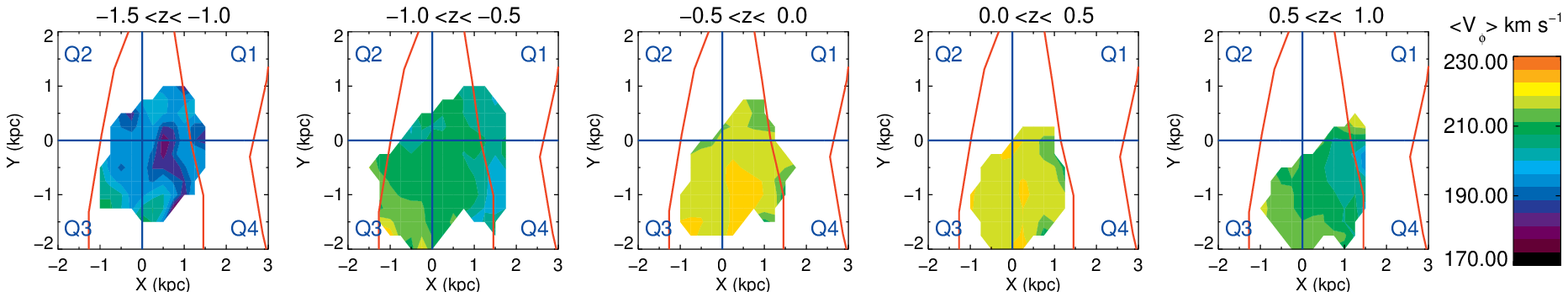} \\
\includegraphics[width=17cm]{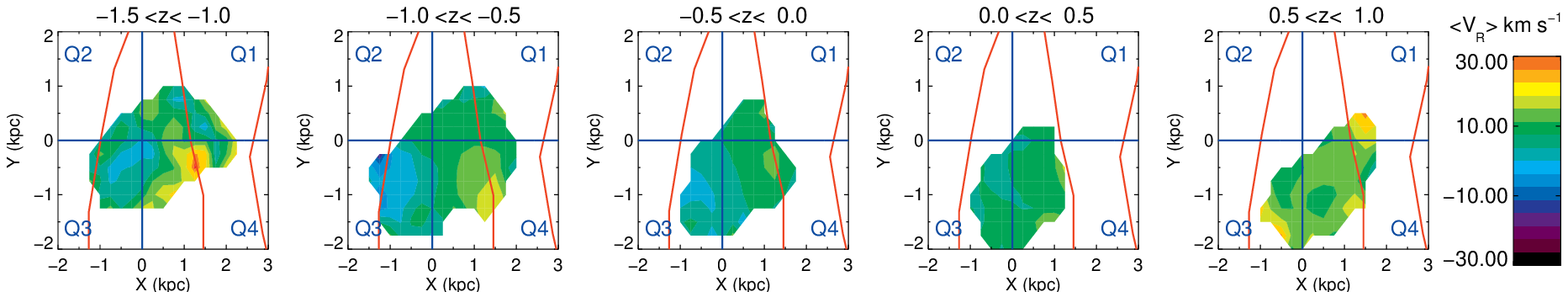} \\
\includegraphics[width=17cm]{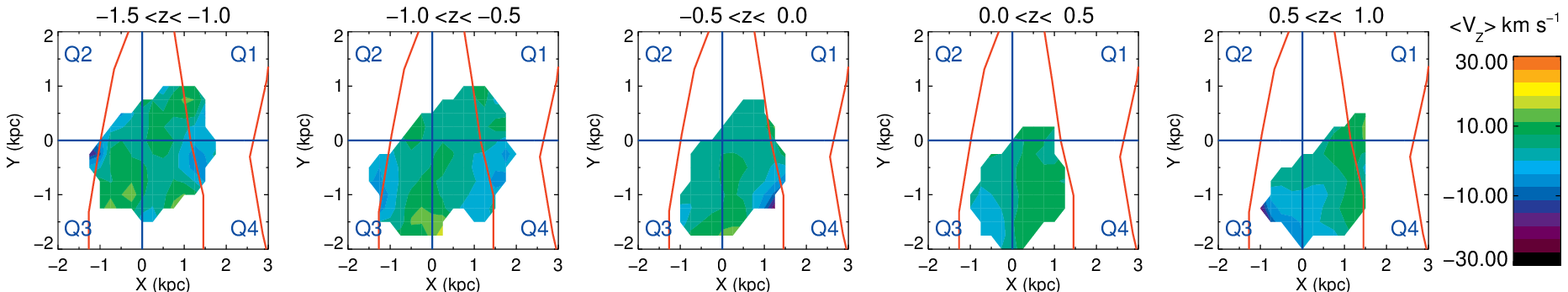} \\
\end{tabular}
\caption{The trends in average $\VPHI$ (top), $\VR$ (middle) and $\VZ$ (bottom) as a function of ($X,\, Y,\, Z$) position in the Galaxy for RC stars only using RAVE catalog proper motions. The position of the spiral arms (red lines) are also plotted. }
\label{fig14}
\end{figure*}

\subsection{$\VPHI$}
\label{sec6.2}

The top two panels of Figure \ref{fig10} show that, as we might
expect, $\VPHI$ is largest in the plane and decreases fairly symmetrically
both above and below the plane. At given values of $Z$, $\VPHI$ also
decreases inwards. These trends are independent of the distances used (RC or
Zwitter) and of the adopted proper-motion catalog, and they are exactly what
the Jeans equations lead us to expect: $\VPHI$ decreases as the asymmetric
drift increases and the asymmetric drift increases with the velocity
dispersion \citep[see e.g.][eq.~(4.34)]{Binney}. In Figure \ref{fig11} the velocity dispersion trends in $\VPHI$ are shown for the RAVE results and the \textsc{Galaxia} model using the RC-assumption distances. The increase of velocity dispersion with increasing $Z$ and decreasing $R$ is seen in both the data and the model. Consequently, whether we
move inwards or away from the plane to regions of increased velocity
dispersion, $\VPHI$ will decrease.
 
Figure \ref{fig8} shows that adopting a different proper-motion catalog does slightly change the predicted values of $\VPHI$: away from the plane
slightly higher values of order $\Delta\simeq5-10\,\kms$ are obtained with the SPM4 proper motions than those
in the RAVE catalog, while the UCAC3 proper motions give intermediate values ($\Delta\simeq2-5\,\kms$).
However, these differences between results from various proper-motion
catalogs are smaller than the difference between the observations and
the predictions of \textsc{Galaxia}: the latter predicts a markedly flatter
profile of $\VPHI$ with $Z$. In other words, the data imply that $\VPHI$
falls away as we move away from the plane significantly more rapidly than
\textsc{Galaxia} predicts. The data give values of $\VPHI$ $5\,\kms$ greater than \textsc{Galaxia} at $Z=0\,\kpc$, falling below the model at $|Z|=0.5\,\kpc$ to be $-10\,\kms$ lower at $|Z|=1\kpc$.
 
In Section \ref{sec4.1.1} we saw that the assumption of a single RC magnitude
can lead to an under-estimation of $\VPHI$ by $\sim5\,\kms$. This effect is also evident in
Figure \ref{fig8} in that the dotted green lines obtained for the \textsc{Galaxia} model using RC
distances lie below the dashed blue lines obtained with the true distances, again by about $5\,\kms$. Further, in Section \ref{sec4.1.2} and Figure \ref{fig5} we saw that for large values of $Z$, $M_K$ normalizations B and C give lower values of $\VPHI$ as large as $15\,\kms$. The effect was only at the extremities of the data however. 
Since even the green lines lie above the data for $|Z|>0.5\,\kpc$, we
conclude that the use of a single RC magnitude or the chosen $M_K$ normalization does not explain the offset
between the data and the predictions of \textsc{Galaxia} and the Galaxy's
velocity field must differ materially from that input into \textsc{Galaxia}.
Comparison of the top and bottom panels of Figures \ref{fig10} reveals that, compared to the predictions of \textsc{Galaxia}, the observations show more clearly 
the expected tendency for $\VPHI$ to decrease as we move in at fixed $|Z|$.

In their study of the HTDC, \citealt{Parker2004} measured the line-of-sight velocities of thick-disk and
halo stars, finding that in Quadrant 1 the rotation of this body of stars
lagged the LSR by 80-90$\,\kms$, while in Quadrant 4 the corresponding lag
was only $20\kms$. Far from confirming this effect, the top-left panel of
Figure \ref{fig14} suggests if anything the opposite is true: at
$X\simeq1.5\kpc$ and $-1<Z<-0.5\,\kpc$, $\VPHI$ is lower in the Quadrant 4
than Quadrant 1. 

The red curves of Figure \ref{fig14} show the positions of spiral arms.
Close to the plane, at $-0.5<Z<0.5$ there is some hint that there is an
increasing lag in $\VPHI$ associated with the spiral features. Further from
the plane, such an association is less clear, as one might expect.

The $(R,\,Z)$ dependence of $\VPHI$ is best described by the power-law:
\begin{equation}
\VPHI=a_1 +\left(a_2+a_3 \frac{R}{\kpc}\right)\left|{\frac{Z}{\kpc}}\right|^{a_4}\,\kms
\label{eqn_vphi}
\end{equation}
with Table \ref{tab4} giving the coefficients for the three different proper motion sources. 

The form is not dissimilar to that given for high-metallicity stars in \citealt{Ivezic2008} between $2<R<16\,\kpc$ and $0<|Z|<9\,\kpc$ ($\langle V_Y\rangle=20.1+19.2|Z/\kpc|^{1.25}\, \kms$) and \citealt{Bond2010} between $0<R<20\,\kpc$ and $0<|Z|<10\,\kpc$ ($\langle \VPHI \rangle=-205+19.2|Z/kpc|^{1.25}\,\kms$). These results are not directly comparable to our fitting formula as we have an extra component that takes radial $R$ dependence into account, which may lead to interaction between the fit coefficients. Nevertheless, our range of values for the exponent (1.06-1.38) for the vertical $Z$ dependence is similar to their results (1.25).

 \begin{table*}
\begin{footnotesize}
\begin{tabular}{lrrrrrr}
\hline
&$a_1$&$a_2$&$a_3$&$a_4$&$Z \,(\kpc)$&$R \,(\kpc)$\\
 \hline
RAVE catalog&225&-51.2&2.6&1.06&$[-1.5,\ 1.5]$&$[6,\ 9]$\\
SPM4&222&-40.2&2.1&1.38&$[-1.5\ 1.0]$&$[7,\ 9]$\\
UCAC3&224&-54.7&3.2&1.10&$[-1.5,\ 1.0]$&$[7,\ 9]$\\
  \end{tabular}
\caption{Parameters for the fit given by Equation \ref{eqn_vphi} to the $\VPHI$ trends using the three proper motion sources; the RAVE-catalog compiled results, SPM4 and UCAC3. The ranges of validity in $R$ and $Z$ are also given.}
\label{tab4}
\end{footnotesize}
\end{table*}

\subsection{$\VR$}
\label{sec6.1}
In the top panels of Figure \ref{fig12}, contours of constant $\VR$ are by no
means vertical, whether one adopts the proper motions in the RAVE or SPM4
catalogs. In fact, with the SPM4 proper motions the contours are not far from
horizontal. Thus the radial gradient $\delta\VR/\delta R=-3\kms/\kpc$
reported by S11 is at the very least just one aspect of a complex
phenomenon. In fact, if the SPM4 proper motions are correct, the trend in
$\VR$ is essentially that the further stars are from the plane, the more they
are moving away from the Galactic centre.

The middle panel of Figure \ref{fig8} shows this situation in a different way by showing that the
red curves for the SPM4 data are horizontal within the errors at all
distances from the plane, but they move down from $\VR\simeq8\kms$ at
$Z\simeq-1.25\,\kpc$ to zero in the plane and back up to $\sim8\kms$ at
$Z\sim0.75\,\kpc$. In this figure the black curves that join the RAVE
data-points tell a different story in that in the panels for $Z\la0$ they
slope firmly downwards to the right. In the two panels for $Z>0$ the data
points from the RAVE proper motions are consistent with no trend in $\VR$
with $R$ with the exception of the innermost point in the panel for
$Z\sim0.25\,\kpc$, which carries a large errorbar. The present analysis is
consistent with the value of $\delta\VR/\delta R$ reported by S11 in that
$-3\kms/\kpc$ is roughly the average of the gradient $\delta\VR/\delta
R\simeq-7\:\mathrm{to}\:-8\kms/\kpc$ given by the RAVE proper motions at $Z<0$ and the
vanishing gradient at $Z>0$.

If the dominant gradient in $\VR$ is essentially in the vertical direction
and an even function of $Z$ as the SPM4 proper motions imply, the suspicion arises that it is an artifact generated by the clear, and expected,
gradient of the same type that we see in $\VPHI$. The gradient could be then seen to be caused by systematics in the proper motions creating a correlation between the measured value of $\VR$ and $\VPHI$ (see e.g. \citealt{Schoenrich2012}). In Section \ref{sec7} we re-explore the line-of-sight detection of the $\VR$, which is a proper-motion-free approach to observing the radial gradient, which however corroborates the existence of a gradient and North-South differences. 

\citealt{Schoenrich2012b} found a larger value of $U_\odot=14\,\kms$ compared to that used in this study and S11, and suggest that this larger value would reduce the gradient in $\VR$. In Figure \ref{fig9} we plot $\VR^s$, the Galactocentric radial velocity with $U_\odot=14\,\kms$. The qualitative trends are unaffected by the higher value of $U_\odot$, with the values only shifted down, bringing the values in the $0.5<Z<1.0\,\kpc$ bin closer to those predicted by \textsc{Galaxia}.

In Figure \ref{fig14} the location of the spiral arms is overlaid for reference on the full $XYZ$ plots of $\VR$ trends. There is some indication that structure in $\VR$ is coincident with spiral arms in the $-1.0<Z<-0.5\,\kpc$ region, with Quadrant 3 and Quadrant 4 showing the largest gradient. With a scale height of the thin disk of 300 pc \citep{Gilmore1983}, at these large $Z$ values we would not expect however that spiral resonances would play a role. However, the recent results of S12 which explained the gradient in terms of such resonances suggests that they do exert an influence at large distances from the Galactic plane. 
The diminishment of the gradient at positive $Z$ values is also evident in this figure. Further modelling in 3D is required to understand if the North-South differences in the gradient can be explained in terms of resonances due to spiral structure, which one would expect to produce effects symmetric with respect to $Z$.

The bottom panels of Figure \ref{fig12} shows the corresponding \textsc{Galaxia} model for the red clump. Here we see that there is an absence of the observed North-South asymmetry, and further, we confirm that the asymmetry is not produced by using the red clump distance estimate. We have explored the influence of changing the distances using the Zwitter distances. We found that these alternative distance calculations also show the same overall $R,\ Z$ variations for $\VR$, while exhibiting some differences in various regions. This indicates that while the distances may influence the quantitative results, qualitatively the overall trends remain the same. 

\subsection{$\VZ$}
\label{sec6.3}

The top panels of Figure \ref{fig13} show that both proper-motion
catalogs yield a similar dependence of $\VZ$ on $(R,Z)$. We see a ridge of
enhanced $\VZ$ that slopes across the panel, making an angle of approximately
$40^\circ$ with the $R$ axis and cutting that axis at approximately the solar radius. This feature is most pronounced in the bottom-left panel, for proper
motions in the RAVE catalog. If we assume the LSR $\VZ$ value is zero, and thus the top of the ridge has $\VZ=0$, we could interpret the stars interior to this point showing an overall rarefaction: stars below the plane move downwards, while those above the plane, move upwards. Exterior to $R=8\,\kpc$ the behaviour is reversed and shows a compression; below the plane stars move upwards, while above the plane stars move downwards. The amplitude of these variations is large; up to $|\VZ|=17\,\kms$ as seen from Figure \ref{fig8}. 

The results for the SPM4 and UCAC3 proper motions do not exhibit quite so large an amplitude in the $\VZ$ variations, but as shown in Figure \ref{fig13} for SPM4, exhibit  a similar overall behaviour, with the line of higher $\VZ$ running diagonal across the $RZ$ plane. However, in Figure \ref{fig8} we see that below the plane, the alternative proper motion results do not show the large dip in $\VZ$ at $R=9\,\kpc$, though this is mostly due to the smaller sampling area in the $XY$ plane of the alternative proper motion sources. Above the plane, the behaviour of the results from the various proper motion sources is similar, with clear deviation from the \textsc{Galaxia} model's predictions in the bottom panels of Figure \ref{fig13}. The \textsc{Galaxia} model itself shows a random pattern of high and low $\VZ$ of the order of $5\,\kms$. There is some smoothing spatially of these fluctuations with the use of RC-magnitude derived distances, though the observed pattern is not generated with this distance method. In general, the proper motions dominate the calculated $\VZ$ values, and the differences in observed structure reflect this. Nonetheless, it is reassuring that the overall pattern is recovered in Figure \ref{fig13} with each proper motion catalog source.

Recently, \citealt{Widrow2012} found with main-sequence SDSS stars that the vertical number-density and $\VZ$ profiles suggest vertical waves in the Galactic disk excited by a recent perturbation. A direct comparison with their results and ours is not possible as their stars are outside the RAVE spatial sampling region. However, our results support this proposition: the rarefaction and compression behaviour seen in the top panels of Figure \ref{fig13} is indicative of wave-like behaviour. This could then be seen as further ``ringing" behaviour of the disk caused by a recent accretion event, as in \citealt{Minchev2009, Gomez2012}. Indeed, \citealt{Gomez2013} further suggest that such vertical waves may have been excited by the Sagittarius dwarf as it passed through the disk. Their simulations find deviations of up to $8\,\kms$ in $\VZ$ in a heavy-Sgr scenario.

Turning now to Figure \ref{fig14} we find some suggestion that the $\VZ$ field shows some alignment with spiral-arms feature, even at significant distances from the plane. Below the plane, the Sagittarius-Carina arm at $X\sim1.5\,\kpc$ is aligned with the low $\VZ$ velocities seen with the RAVE catalog proper motions for $Y<0$. In this representation, the SPM4 and UCAC3 proper motion results also show a dip in their $\VZ$ velocities around this area, though the feature is not as pronounced as for the RAVE catalog proper motions. The Perseus arm is also aligned with an area of lower $\VZ$ at $X=-1.25,\ Y<0,\ Z< -0.5\,\kpc$. Across all values of $Z$, the alignment of the $\VZ$ features with the spiral arms is less clear for $Y>0$, which below the plane may be explained by the proximity of the Hercules thick disk cloud.\citealt{Siebert2012} showed how spiral density waves could be used to explain the $\VR$ streaming motion. Siebert et al. (in preparation) goes further to investigate how they could also create the $\VZ$ structures, coupled to $\VR$.

These results suggest that there are large-scale vertical movements of stars, with various cohorts at the same $Z$ moving in opposing directions. In \citealt{Casetti2011}, it was proposed that while the Galactic warp starts at the solar radius, the small elevations of the warp at these distances ($70-200\,\textrm{pc}$ for $R=8-10\,\kpc$ \citep{Lopez2002}) and the large distances of the RAVE stars from the plane would mean that kinematics of stars would be little affected by the warp. In \citealt{Russiel2002}, the nearby Sagittarius-Carina arm as traced by star-forming complexes is found to lie mostly below the plane by $100-200\pc$, which suggests that nearby spiral arms can be influenced by some warping. Whether the warp is long-lived or a transient feature would have implications for the associated velocities. The complexity of the vertical velocity distribution suggested by our results would tend to point towards transient features: they suggest a non-equilibrium state. A multi-mode travelling wave caused by a recent accretion event, or structure associated with the disk's spiral arms, would be the most likely explanations for the observed velocity structure.

\subsection{Velocity dispersion trends}
\label{sec6.4}

The second moments of the velocity components increase with increasing $Z$ and decreasing $R$ in a smooth fashion as expected, with Figure \ref{fig11} showing the results for $\sigma(\VPHI)$. As for $\VPHI$, they can be fit by a simple parametric function.

\begin{figure*}
\begin{center}
\includegraphics[width=1\linewidth]{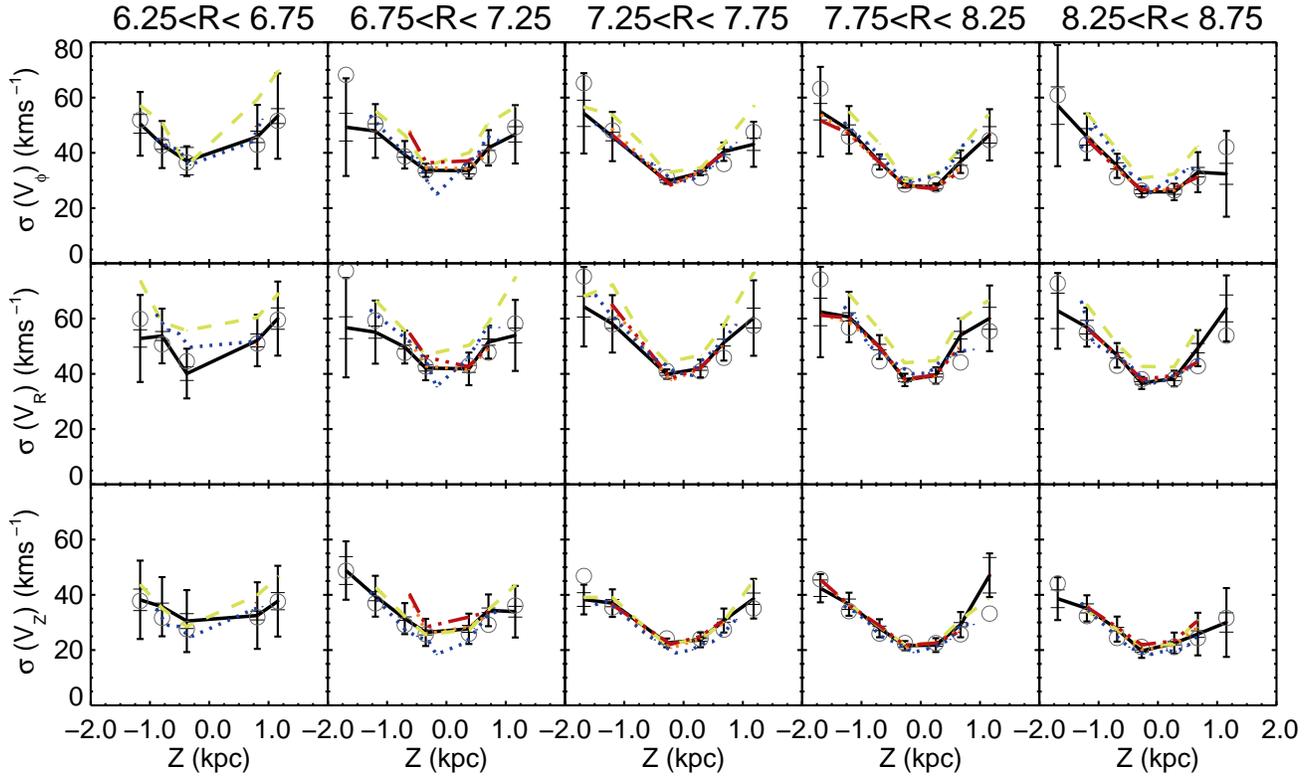}
\caption{The trends in velocity dispersion as a function of ($R,\ Z$) position in the Galaxy for RC stars with RAVE catalog proper motions (black solid line), UCAC3 proper motions (orange dashed line) and SPM4 proper motions (red dot-dashed line). \textsc{Galaxia} model results are also given, using the real \textsc{Galaxia} distances (dark blue) and RC distances (light green). Error bars give the measurement (thick line, short hat) and Poisson (thin line, long hat) errors. The open circles give the results of the fitting formula using the coefficients from the RAVE catalog proper motions fit.}
\label{fig15}
\end{center}
\end{figure*}

Figure \ref{fig15} displays the trends in the dispersion of $\VPHI,\ \VR,\ \VZ$ as a function of $Z$ for $0.5\,\kpc$ thick slices in $R$ using $0.5\,\kpc$ bins in $Z$. Also plotted are the results based on UCAC3 and SPM4 proper motions, and the \textsc{Galaxia} model results using the ``true" model distances plus RC-calculated distances. Unlike the mean-value trends, the differences between the various proper motion sources is minimal. We see however that the \textsc{Galaxia} model results indicate that the use of RC-calculated distances can lead to an increase in the measured dispersion values, particularly for $\VR$. Interestingly however, there is a reasonable agreement between the results of \textsc{Galaxia} and that observed with RAVE. 

We note that the dispersions were corrected for errors following the same methodology of \citealt{Casetti2011}. This method adjusts a guess of the true dispersion, $\sigma_V$, in velocity component $V$ until
\begin{equation}
\mathrm{stddev}(({V_i-\bar{V}})/{\sigma_T}) = 1
\end{equation}
where $\sigma_T^2=\sigma_V^2+eV_i^2$, with $eV_i$ the MC-derived error in $V$ for $i$th star and $\bar{V}=\mathrm{mean}(V_i)$ is the mean over the sample. 
 
The velocity dispersions exhibit linear behaviour in $R$ and parabolic behaviour in $Z$. Thus, a fit to the dispersion trends was found to be provided by the simple form 
\begin{equation}
\sigma_V=b_1+b_2 \frac{R}{\kpc} + b_3 \frac{Z}{\kpc}^2\,\kms
\label{eqn_disp}
\end{equation}
with Table \ref{tab5} giving the coefficients for the three velocity components for the three different proper motion sources. The similar coefficients for each velocity component underlines the similarity of the trends between the three sources.

 \begin{table}
\begin{footnotesize}
\begin{tabular}{lrrr}
\hline
&$b_1$&$b_2$&$b_3$\\
 \hline
$\VPHI$&&&\\
RAVE catalog&67.4&-5.0&12.4\\
SPM4&75.7&-6.1&15.8\\
UCAC3&71.2&-5.5&15.2\\
 \hline
 $\VR$&&&\\
RAVE catalog&62.6&-3.0&12.5\\
SPM4&60.0&-2.7&15.1\\
UCAC3&63.3&-3.1&13.5\\
 \hline
 $\VZ$&&&\\
RAVE catalog&47.0&-3.1&8.2\\
SPM4&49.4&-3.4&9.8\\
UCAC3&43.6&-2.8&10.4\\
\end{tabular}
\caption{Parameters for the fit given by Equation \ref{eqn_disp} to the velocity dispersion trends using the three proper motion sources; the RAVE-catalog compiled results, SPM4 and UCAC3. The parameters are valid for the same ranges given in Table \ref{tab4}.}
\label{tab5}
\end{footnotesize}
\end{table}

\section{Velocity gradient: line-of-sight detection}
\label{sec7}

In S11 the gradient in $\VR$ was first investigated using the line-of-sight velocity, $\vlos$, along narrow lines towards the centre and anti-centre. The rationale for this approach was to observe the $\VR$ gradient independent of the proper motions. The complicated 3D structure of the Galactic $\VR$ profile was not considered so we revisit this approach, keeping in mind the 3D trends.

\subsection{$\VR$}
\label{sec7.1}

\begin{figure}
\begin{center}
\includegraphics[width=\linewidth]{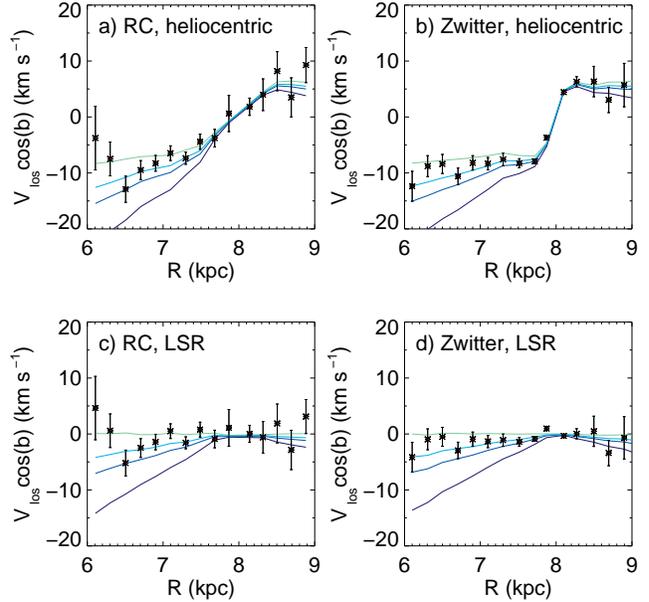}
\caption{As in Figure 3 of S11, projection of mean RAVE $\RV$ on the Galactic plane in distance intervals of $200\,\pc$ towards the Galactic centre ($|l|<5\deg$) and anti-centre ($175\deg<l<185\deg$). Panels (a) and (c) show the results for RC stars with heliocentric line-of-sight velocity and corrected to the LSR, respectively, while (b) and (d) similarly show the Zwitter distance results. The solid curves represent a thin disk population with a radial velocity gradient of $\delta \VR/\delta R= 0,\ -3,\ -5$ and $-10\ \kms/\kpc$, going from green to purple.}
\label{fig16}
\end{center}
\end{figure}

\begin{figure}
\begin{center}
\includegraphics[width=\linewidth]{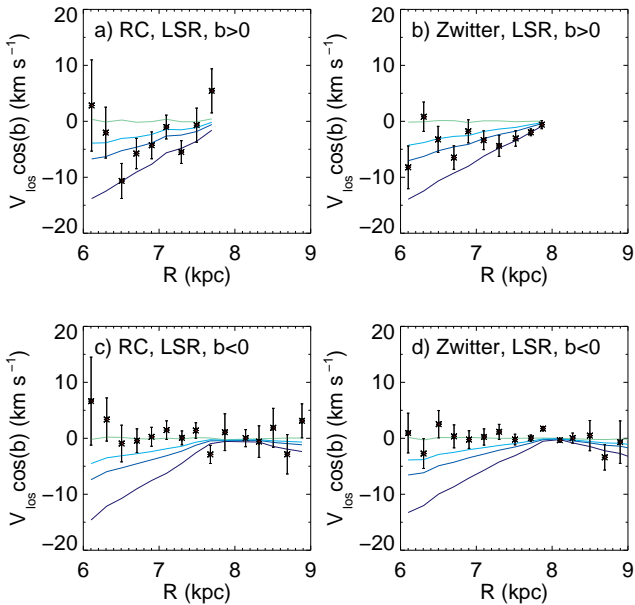}
\caption{As in the bottom plots of Figure \ref{fig16}, but for Galactic centre ($|l|<7\deg$) and anti-centre ($173\deg<l<187\deg$) and split into above (a, b) and below (c, d) the plane.}
\label{fig17}
\end{center}
\end{figure}

\begin{figure}
\begin{center}
\includegraphics[width=\linewidth]{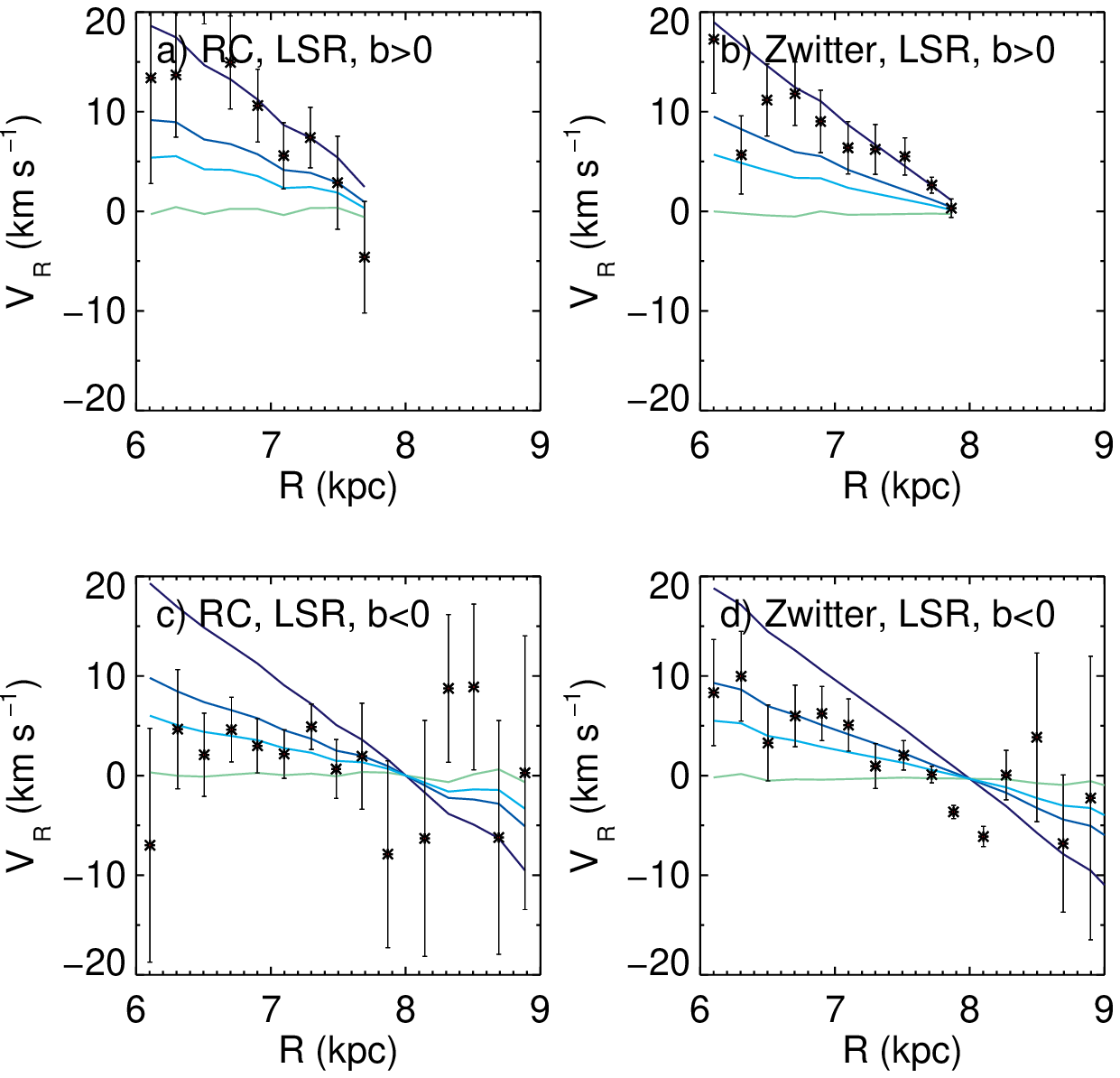}
\caption{Corresponding $\VR$ trends to Figure \ref{fig17}, looking at the Galactic centre ($|l|<7\deg$) and anti-centre ($173\deg<l<187\deg$) and split into above (a, b) and below (c, d) the plane.}
\label{fig18}
\end{center}
\end{figure}

\begin{figure}
\begin{center}
\includegraphics[width=\linewidth]{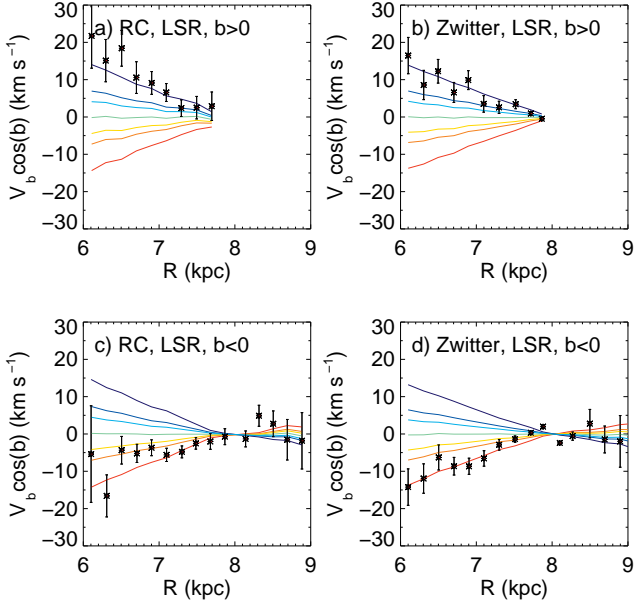}
\caption{$V'_b \cos{b}$ for Galactic centre ($|l|<7\deg$) and anti-centre ($173\deg<l<187\deg$) directions and split into above (a, b) and below (c, d) the plane. The solid curves represent a thin disk population with a radial velocity gradient of $\delta \VZ/\delta R=10,\ 5,\ 3,\ 0,\ -3,\ -5$ and $-10\ \kms/\kpc$, going from red to green to purple.}
\label{fig19}
\end{center}
\end{figure}

\begin{figure}
\begin{center}
\includegraphics[width=\linewidth]{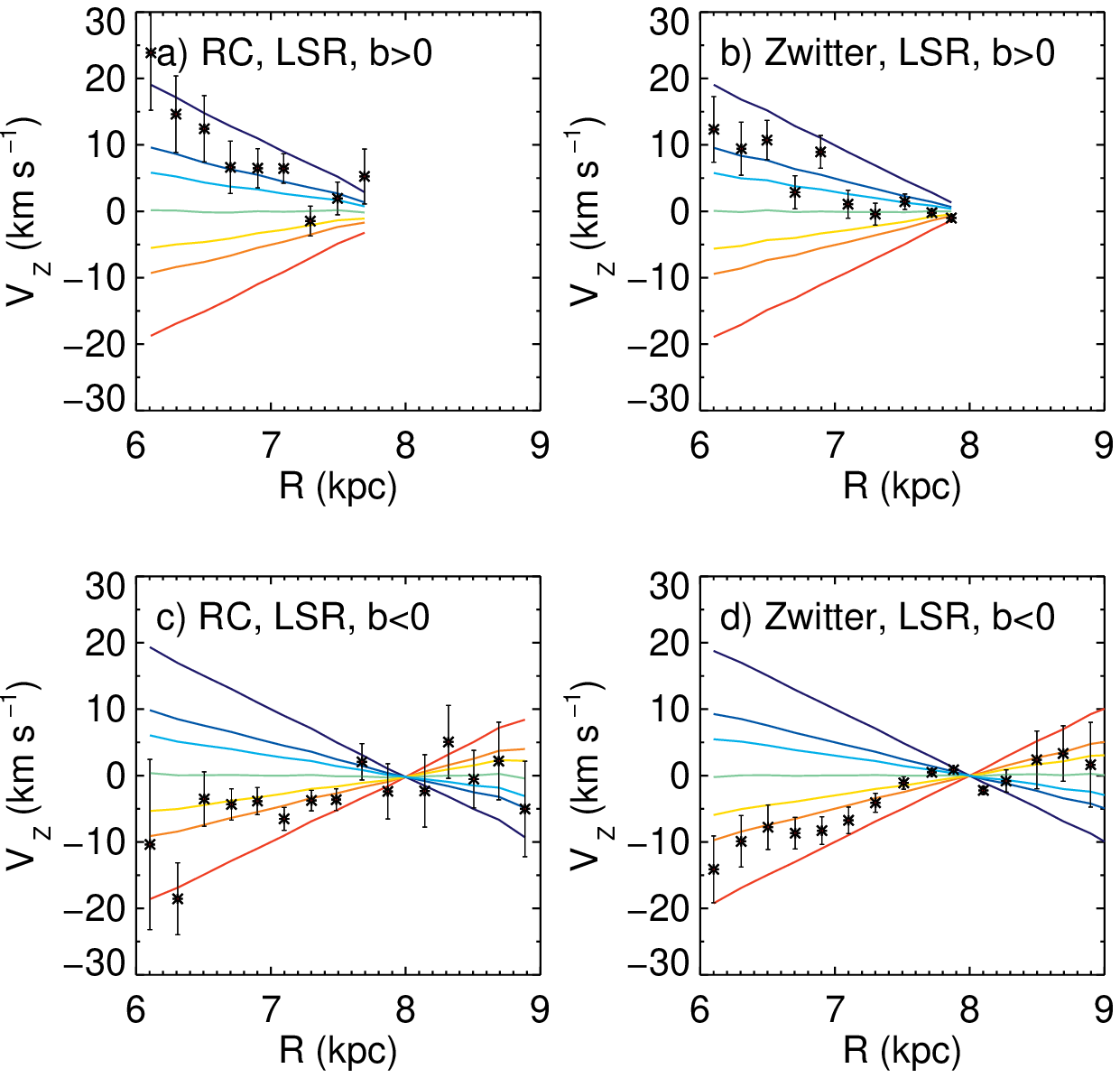}
\caption{Corresponding $\VZ$ trends to Figure \ref{fig19}, looking at the Galactic centre ($|l|<7\deg$) and anti-centre ($173\deg<l<187\deg$) directions and split into above (a, b) and below (c, d) the plane.}
\label{fig20}
\end{center}
\end{figure}

Figure 3 from S11 gives the projection onto the plane of the mean $\vlos$ in bins $200\,\pc$ wide as a function of $d\cos{l}\cos{b}$, with the latter a proxy for $X$ in their Galactic co-ordinates, with $X_\textrm{S11}\sim R_\odot - d\cos{l}\cos{b}$. For $|l|<5\deg$ and $175\deg<l<185\deg$ the values of the radial component, $\VR$ is essentially the same as the Cartesian value $-U$. As most stars in these narrow cones have $\cos{b}/\sin{b}\sim 1.5$, $U$ (and so $-\VR$) is dominated by the term $\vlos \cos{b}$. In the Appendix we list the relevant equations which shows how this follows. S11b compared the mean velocities to those expected for a thin disk in circular rotation and adding a radial gradient of $\partial(\VR)/\partial R=-3, -5$ and $-10\ \kms/\kpc$ (plus a thick disk lagging the LSR by $36\,\kms$), finding that the results were consistent with a radial gradient of  $\partial(\VR)/\partial R=-3\ \mathrm{to} -5\,\kms/\kpc$. 

From Equations 1 and 7 in the Appendix, we see that if we are to consider the $\vlos$ as representative of $U$ (and so $-\VR$), technically it would be better to first subtract the solar motion from the line-of-sight velocity, to obtain $\vlos'$. Otherwise the components above and below the plane will be shifted from each other, which can be problematic if the sample shifts from above-the-plane to below-the-plane, potentially producing spurious trends. This was omitted in the analysis of Figure 3 from S11, however, it did not affect the conclusions as the models were similarly shifted. 

In Figure \ref{fig16} we re-examine the trends in $\vlos \cos{b}$ for the RC giants and stars with Zwitter distances, where we plot against $R$ rather than $d\cos{l}cos{b}$ as this is easier to interpret and there is little difference between the two values. We plot the results with and without the LSR correction, with the curve being flattened in the latter with the removal of the $U_\odot$ component. We also plot the trends for a thin-disk as in S11, using 100 Monte Carlo realizations. We see here that not correcting for the solar motion does not affect the conclusions. Note that with the updated data sets the $\VR$ streaming motion is slightly less apparent in both the RC and Zwitter results than it was in S11: the model with $\partial(\VR)/\partial R=-3\,\kms/\kpc$ is a better fit to the data. Also, the differences between the RC and Zwitter trends around the solar radius in Figure \ref{fig16} is a result of the different geometric sampling of the two. 

In Figure \ref{fig17} we examine the $\vlos' \cos{b}$ results split into North and South components. In this we open up the criteria a bit to within $7\deg$ of the centre and anti-centre direction to reduce the effects of Poisson noise and make the trends clearer. Here we see that that differences can be discerned between the North and South trends, however, they are opposite to that seen in the previous section; the gradient is in the North ($-5\ \mathrm{to} -10\,\kms/\kpc$) is stronger though absent in the South ($0\,\kms/\kpc$)) for $R<8\,\kpc$. Plotting the corresponding results for $\VR$ in Figure \ref{fig18}, we find that the including the proper motion results shifts the overall values so that the northern trends are more in line with $\delta \VZ/\delta R=-10\,\kms/\kpc$ and the southern $\delta \VZ/\delta R=-3\ \mathrm{to} -5\,\kms/\kpc$. Despite the disparity with the actual numbers, the cause of which is discussed below, both $\vlos' \cos{b}$ and the $\VR$ nonetheless exhibit differences between the north and south trends, opposite to those found in Section \ref{sec6.1}.

To understand the reversal of the trends note first that the selected sample cuts across a range of $Z$ as we change $R$ so these plots combine the $R$ and $Z$ trends seen previously. Second, the necessary restriction on $l$ means that we are sampling a very narrow beam and thus do not see the global patterns, but only those along that beam. In Figure \ref{fig14} Quadrants 3 and 4 show the largest gradient, which we miss with the beams. Hence, these plots emphasize the 3D nature of the $\VR$ values in the solar neighbourhood: what you measure very much depends on where you look, be it north or south, and at different $R$ and $Z$ values. 

\subsection{$\VZ$}
\label{sec7.2}

From Equation 6 in the Appendix we can see that $V'_b \cos{b}$ can be used as a proxy for $W=\VZ$ if $|\cos{b}|>|\sin{b}|$. This condition is for the most part met in the low latitudes sampled along the $|l|<7\deg$ and $173\deg<l<187\deg$ cones used above. So in Figure \ref{fig19} we examine the trends above and below the plane for $V'_b \cos{b}$ in these directions, with Figure \ref{fig20} giving the corresponding trends in the total $\VZ$. Note that we use RC and Zwitter distances here to provide the abscissa, supplementing the proper motion data. We also plot both positive and negative trends in $\VZ$ for a thin disk model as above, with values from $\partial(\VZ)/\partial R=-10, -5, -3, 0, 3, 5, 10\ \kms/\kpc$. 

Both Figures \ref{fig19} and \ref{fig20} show that the trends above and below the plane are markedly different, with the $\VZ$ decreasing with R above the plane at a rate of $\partial(\VZ)/\partial R \sim -10\ \kms/\kpc$ according to the $V'_b \cos{b}$ plot and $-5\ \kms/\kpc$, to the $\VZ$ plot. Below the plane, there is a positive trend of $\partial(\VZ)/\partial R \sim +5\,\kms/\kpc$ in both plots. This positive trend in $\VZ$ appears to stop just beyond the solar circle at $R\sim8.5\,\kpc$. In contrast to $\VR$ above, this behaviour is consistent with what was found in Section \ref{sec6.3}. 

Note that the differences between the observed magnitude of the trends in $\vlos' \cos{b}$ and $\VR$, plus $V'_b \cos{b}$ and $\VZ$ can be explained by the fact that both $\vlos'$ and $V'_b$ contain components of $U$ and $W$. For small $l$, $\vlos' \cos{b}=U'\cos^2{b}+W'\cos{b}\sin{b}$ and $V'_b \cos{b}=-U'\sin{b}\cos{b}+W'\cos^2{b}$. The cross-terms mean that there is some `leakage' of the $\VZ$ trends into $\vlos' \cos{b}$ and $\VR$ trends into $V'_b \cos{b}$, which work to either diminish or enhance the observed trend in the single component. Nonetheless, the fact that there \textit{are} differences between the north and south for the single components at all is further evidence of the 3D variations of the velocity values.

The $\VR$ and $\VZ$ trends, as seen via the line-of-sight velocities and proper motions respectively, are unaffected by potential systematics in the distances: a change of the distance scale would not affect the fact that a trend is seen at all. This is particularly so for the differences observed between the northern and southern samples. Furthermore, both the SPM and UCAC3 proper motions give similar results as in Figures \ref{fig19}, with a stark negative trend above the plane and a positive trend below for the region $R<8.5\,\kpc$. Thus, neither the distances nor the proper motions introduce large systematics into this method of detection. 

\section{Conclusion}
\label{sec8}

Using RAVE red clump giants we have examined in detail the first moments of the velocity components in a large volume around the Sun. We find differences between the North and South in the streaming motion reported in \citealt{Siebert2011} in Galactocentric radial velocity, $\VR$. Above the plane, there is a large outward flow ($\VR=8-10\,\kms$ for $0<Z<1\,\kpc$) with a shallow or non-existent gradient. Below the plane, there are lower values of $\VR$ outside the solar circle, down to $\VR=-10\,\kms$ at $R=9\,\kpc,\:-1<Z<-0.5\,\kpc$. This is associated with a much steeper gradient in $R$, particularly in Quadrants 3 and 4, the largest gradient being $\delta\VR/\delta R
=-8\,\kms/\kpc$ for $-1<Z<-0.5\,\kpc$.

The behaviour of $\VZ$ shows a surprising complexity suggestive of a wave of compression and rarefaction: there is a ridge of higher $\VZ$ passing at an angle of $40^\circ$ to the plane, intersecting the plane at roughly the solar radius. Assuming the LSR $\VZ$ to be zero, stars interior to the solar circle and above the plane are moving upwards, while those below the plane, downwards. Exterior to the solar circle, stars both above and below are moving on average towards the plane. Values of up to $|\VZ|=17\,\kms$ are observed. We confirm these differences by examining the transverse velocities along narrow cones towards- and away-from the Galactic centre. 

Physically, the $\VZ$ velocity field implies alternate rarefaction and
compression, as in a sound wave. Thus our three-dimensional velocity field
confirms the recent one-dimensional results of \citealt{Widrow2012} in which
North-South differences in the velocities of SDSS stars suggested vertical
waves in the Galactic disk. Two likely causes of these waves are either a recently accreted satellite or the disk's spiral arms. Further modelling will hopefully help decide between these two scenarios.

$\VPHI$ is much more regular than the other two components, showing the most qualitative agreement with the mock sample created with the \textsc{Galaxia} model for the Galaxy and the expected increase of $\VPHI$ with increasing $Z$ and decreasing $R$. The model gives a much flatter profile of $\VPHI$ with $Z$ however than the data; at $Z=0\,\kpc$ we measure $\VPHI$ values $5\,\kms$ larger than \textsc{Galaxia}, falling to $-10\,\kms$ lower at $|Z|=-1\,\kpc$. There is some hint of an increased lag in $\VPHI$ near the plane associated with spiral arm features. We present a simple parametric fit to the $\VPHI$ dependence on $R$ and $Z$.

We also trace the second moments of the velocities as a function of $R$ and $Z$, providing a simple parametric fit to these trends as well. 

The red clump is increasingly being used as a standard candle for field stars, given its ease of identification and the relative insensitivity of the $M_K$ magnitude to age and metallicity. We modelled our selection of RC stars using \textsc{Galaxia}, showing a surprisingly high level of contamination by first-ascent giants despite a tight selection in the $\log g$-$(J-K)$ plane. However, given that the majority of these giants have a similar magnitude to the red clump itself, the effect on the distances does not render them unusable. It means though that there is further complexity in the metallicity-age mixture of selected red clump stars: the population is by no means homogenous in age and abundance. 

The assumption of a single $M_K$ magnitude for the RC and the proper motions are the largest sources of systematic error in our analysis. Indeed, a deeper study into proper motion differences is required to establish which of the catalogs is most trustworthy. Nevertheless, we have established that North-South differences do exist in $\VR$ and $\VZ$ despite these problems. For $\VR$ a line-of-sight detection, which excludes the proper motions, shows gradients above and below the plane despite the pencil beams in this analysis pointing away from the area of the largest gradient in Quadrants 3 and 4. These results particularly illustrate the 3D nature of the velocity field. For $\VZ$, results using the three proper motion sources give the same rarefaction-compression behaviour, albeit with some variation in the details.

The 3D structure in $\VR$ and $\VZ$ presents challenges to future modelling of the Galactic disk under the influence of the bar, spiral features and any other perturbations (be they temporally localised or not). It is not intrinsically clear indeed if the structure in the two are coupled or arise from different physical mechanisms.

\section*{Acknowledgments}

Funding for RAVE has been provided by: the Australian Astronomical
Observatory; the Leibniz-Institut fuer Astrophysik Potsdam (AIP); the
Australian National University; the Australian Research Council; the
French National Research Agency; the German Research Foundation (SPP 1177 and SFB 881); the European Research Council (ERC-StG 240271 Galactica); the Istituto Nazionale di Astrofisica at Padova; The Johns Hopkins University; the National Science Foundation of the USA (AST-0908326); the W. M. Keck foundation; the Macquarie University; the Netherlands Research School for Astronomy; the Natural Sciences and Engineering Research Council of Canada; the Slovenian Research Agency; the Swiss National Science Foundation; the Science \& Technology Facilities Council of the UK; Opticon; Strasbourg Observatory; and the Universities of Groningen, Heidelberg and Sydney. The RAVE web site is at http://www.rave-survey.org.

We are grateful to the referee for a number of helpful suggestions which improved the paper.

\footnotesize{

\footnotesize{

\appendix
\section*{Appendix}
\label{appen}
The peculiar motion of the Sun is given in the Cartesian values $(U_\odot, V_\odot, W_\odot)$. We define the line-of-sight velocity and proper motion vectors in $(l,\ b)$ co-ordinates corrected for reflex of the solar motion:
\begin{flalign}
&\vlos' = \vlos+(U_\odot \cos{l} \cos{b}  + V_\odot \sin{l} \cos{b} + W_\odot \sin{b}),\\
&V'_l =V_l +( -U_\odot \sin {l} + V_\odot \cos {l}),\\
&V'_b = V_b+(-U_\odot \cos {l} \sin {b} - V_\odot \sin{ l} \sin {b} + W_\odot \cos {b}).
\end{flalign}

The reason we do the correction this way round (rather than simply in $(U, V, W)$) is because we wish to see how the line-of-sight velocities and proper motions are constituted. For example, the Sun is moving up towards the Galactic pole. To correct for this in the line-of-sight velocity when looking above and below the plane this contribution from $W_\odot$ is subtracted and added respectively. Hence, the solar motion must be corrected for before examining trends in $\vlos$, especially if averaging over positive and negative $z$. 

The $(U,\ V,\ W)$ components are given by
\begin{flalign}
&U= \vlos' \cos{l} \cos{b} - V'_l \sin{l}  - V'_b\cos{l} \sin{b},\\ 
&V= \vlos' \sin{l} \cos{b} + V'_l \cos{l} - V'_b\sin{l} \sin{b}, \\
&W =\vlos' \sin{b} + V'_b \cos{b},
\end{flalign}
where $U$ is positive towards the Galactic centre. For small $l$ this reduces to
\begin{flalign}
&U = \vlos' \cos{b}  - V'_b\sin{b},\\
&V = V'_l, \\
&W =\vlos' \sin{b} + V'_b \cos{b}.
\end{flalign}
Similar results are obtained near $l=180\deg$, obviously with some sign changes. To convert to cylindrical co-ordinates we firstly define the Cartesian values corrected for the circular velocity at the solar radius;
\begin{flalign}
&V_X = U\\
&V_Y = V+V_{\mathrm{c}, 0} \\
&V_Z =W
\end{flalign}
where we use the nominal value $V_{\mathrm{c}, 0} =220\ \kms$ in the majority of this paper. The cylindrical components are then
\begin{flalign}
&\VR = ((X-R_\odot)V_X+YV_Y)/R\\
&\VPHI= -((X-R_\odot)V_Y-YV_X)/R \\
&\VZ =\VZ,
\end{flalign}
with $R=\sqrt{(X-R_\odot)^2+Y^2}$. For the small $l$ (or $l\sim180\deg$) discussed in Section \ref{sec7} where $y\simeq0$, and for the solar neighbourhood where $X-R_\odot < 0$,  these reduce to $\VR\simeq -U$ and $\VPHI\simeq V+V_{\mathrm{c}, 0}$ and so these values are interchangeable in this regime.

\end{document}